\documentclass{pasj00}
\draft
\SetRunningHead{C.Hikage et al.}
{Minkowski Functionals of SDSS Galaxies}
\Received{2003/04/09}
\Accepted{2003}
\begin{document}
\title{Minkowski Functionals of SDSS galaxies I : \\
Analysis of Excursion Sets}
\author{%
Chiaki \textsc{Hikage}\altaffilmark{1}, 
Jens \textsc{Schmalzing}\altaffilmark{2}, 
Thomas \textsc{Buchert}\altaffilmark{1,2,3}, 
Yasushi \textsc{Suto}\altaffilmark{1}, \\
Issha \textsc{Kayo}\altaffilmark{1}, 
Atsushi \textsc{Taruya}\altaffilmark{1}, 
Michael S. \textsc{Vogeley}\altaffilmark{4}, 
Fiona \textsc{Hoyle}\altaffilmark{4}, \\
J. Richard \textsc{Gott III}\altaffilmark{5},   
Jon \textsc{Brinkmann}\altaffilmark{6},
et al.}
\altaffiltext{1}{Department of Physics, School of Science, 
University of Tokyo, Tokyo 113-0033} 
\altaffiltext{2}{Theoretische Physik, Ludwig--Maximilians--Universit\"at,
\\ Theresienstr. 37, D--80333 M\"unchen, Germany} 
\altaffiltext{3}{Research Center for the Early Universe
(RESCEU), School of Science, \\  University of Tokyo, Tokyo 113--0033} 
\altaffiltext{4}{Department of Physics, Drexel University, 3141
Chestnut Street, Philadelphia, PA 19104, USA}
\altaffiltext{5}{Princeton University Observatory, Peyton Hall,
Princeton, NJ 08544, USA}
\altaffiltext{6}{Apache Point Observatory, P.O.Box 59, Sunspot NM
88349-0059, USA}
\email{hikage@utap.phys.s.u-tokyo.ac.jp, 
jens@theorie.physik.uni-muenchen.de, \\
buchert@theorie.physik.uni-muenchen.de, suto@phys.s.u-tokyo.ac.jp, \\
kayo@utap.phys.s.u-tokyo.ac.jp, ataruya@utap.phys.s.u-tokyo.ac.jp,\\
vogeley@drexel.edu, hoyle@venus.physics.drexel.edu, 
jrg@astro.princeton.edu, jb@apo.nmsu.edu}
\KeyWords{cosmology: large--scale structure of universe --- 
cosmology: observations --- methods: statistical} 
\maketitle
\begin{abstract}
  We present a first morphometric investigation of a preliminary
  sample from the SDSS of $154287$ galaxies with apparent magnitude
  $14.5<m_{\rm r}<17.5$ and redshift $0.001<z<0.4$.  We measure the
  Minkowski Functionals, which are a complete set of morphological
  descriptors. To account for the complicated wedge--like geometry of
  the present survey data, we construct isodensity contour surfaces
  from the galaxy positions in redshift space and employ two
  complementary methods of computing the Minkowski Functionals.  We
  find that the observed Minkowski Functionals for SDSS galaxies are
  consistent with the prediction of a $\Lambda$--dominated
  spatially--flat Cold Dark Matter model with random--Gaussian initial
  conditions, within the cosmic variance estimated from the
  corresponding mock catalogue. We expect that future releases of the
  SDSS survey will allow us to distinguish morphological differences in
  the galaxy distribution with regard to different morphological type
  and luminosity ranges.
\end{abstract}
\section{Introduction}

The first--year WMAP ({\it Wilkinson Microwave Anisotropy Probe}) data
allow measurement of the parameters of the standard cosmological model
to unprecedented accuracy; it now seems reasonable to assume
$\Omega_{\rm m}\approx0.3$, $\Omega_\Lambda\approx0.7$, and
$h\approx0.7$ (Spergel et al. 2003).  Furthermore, the data put
stringent constraints on non-Gaussian signatures in the primordial
density fluctuations at $z\approx1000$ \citep{K2003,Colley2003}. These
results point to two important and complementary questions for
cosmology in this century: origin and evolution. The former regards
the question of how our universe came to have the observed values of
the cosmological parameters including questions of their intrinsic
nature. The latter question asks how the
universe at $z\approx1000$ revealed by WMAP leads to the rich current
structure traced by galaxies.  The present study is an attempt to
address the latter by characterizing and quantifying the morphological
properties of the galaxy distribution of Sloan Digital Sky Survey
(SDSS) spectroscopic catalogues through estimation of the Minkowski
Functionals (Mecke et al. 1994; Schmalzing and Buchert 1997) .

We have three primary goals for analyzing galaxy catalogues with
Minkowski Functionals (hereafter MFs): the first is to test
Gaussianity of the primordial density fluctuations.  Indeed, tests of
Gaussianity used to be of primary importance before the discovery of
the CMB (Cosmic Microwave Background) temperature anisotropy, but now
have been superseded by the CMB map analysis, although measurement of
the MFs is still useful as an independent tool.  The second goal is to
understand the evolution of galaxy clustering, with particular
emphasis on the morphology of large-scale structure.  Since the
primordial Gaussianity has been fairly well--established, studying the
growth of structure should be the major role of the MFs, which
complement estimation of the N--point correlation functions.  Finally,
we can examine whether the current observational catalogues may be
regarded as fair samples of the density distribution in the Universe.
In reality, however, those primary goals may be achieved barely
partially with the current dataset that we use below. Nevertheless our
results shown below should be regarded as the most careful attempt
towards the goals on the basis of the best observational data
currently available.

Focusing on morphological properties has two major advantages.  First,
there is a firm mathematical basis of morphological statistics
developed in the field of integral geometry, which has condensed the
complex information about morphology that is contained in all orders
of the N--point correlation functions into a set of $d+1$ functionals,
where $d$ is the dimension of the spatial pattern (Mecke et al. 1994;
Kerscher 2000 and references to the mathematical literature therein).
Second, because this reduced set implicitly contains information from
higher--order correlation functions, they provide constraints that are
complementary to low--order correlation function analysis (e.g., Mecke
et al. 1994, Schmalzing et al. 1999b). As a consequence of including
higher--order correlation terms, the MFs also cover
phase--correlations that are essential for the formation of
large--scale structure.  In particular, the MFs are sensitive to
features on spatial scales larger than typical scales
($\approx5h^{-1}$Mpc) accessible by the two--point correlation
function, as demonstrated by Kerscher et al.~1998.

A spatial distribution in three dimensions has four Minkowski
Functionals, namely its volume, its surface area, the integrated mean
curvature and the integrated Gaussian curvature, i.e., the Euler
characteristic.  The latter measures the same property as the Genus (Gott et
al. 1986, Melott 1988) which is a commonly used statistic in
large--scale structure analysis.  Using the whole family of MFs turns
out to provide useful additional information: previous experience
already demonstrated that the total surface area ({\it c.f.,} Ryden
1988; Ryden et al. 1989) and the integral mean curvature are more
significant in discriminating structural differences.  Furthermore,
integral geometry provides robust formulae to deal with bounded data
sets; in fact the boundary effect can only be exactly deconvolved by
using the complete set of MFs (e.g., Kerscher et al. 1998).  Useful
`shapefinders' can be constructed on the basis of isoperimetric ratios
among the MFs (Sahni et al. 1998, Schmalzing et al. 1999a).

A natural method to evaluate the MFs of a point--distribution, such as
a galaxy catalogue, is the so--called `Boolean Grain Model', which
decorates each point in the sample with, in the simplest case,
spherical balls; the body formed by the union of these balls at a
given scale (the radius of the balls) is then measured with MFs (Mecke
et al.\ 1994). This method is usually preferable to the construction
of a density field by smoothing and measurement of the morphology of
isodensity contour surfaces, because the former has only one parameter
(the radius) and does not require a smoothing kernel that discards
information. However, the Boolean Grain Model has limitations when it
comes to analyzing slice--like data such as the present sample of the
SDSS data set: the largest Boolean Grain (sphere) is the one that
fills the smallest extent of the sample and therefore the analysis is
limited to that scale. For this reason we choose to smooth the point
distribution with a Gaussian and construct excursion sets (isodensity
contours) as a sensible way to study the morphology of large--scale
structure in these data. Because the weak point in this analysis is
the reliability of the constructed density field on a given smoothing
scale, we employ two complementary methods of calculating the
MFs of isodensity contour surfaces, which introduces a systematic
control element into our analysis (Schmalzing and Buchert 1997).

\medskip

This paper is organized as follows. In Section 2 we explain and
illustrate the SDSS data set we analyze and describe the mock catalogues
used for comparison. In Section 3 we introduce the Minkowski
Functionals, especially the computational methods of their calculation
(details are provided in Appendix A).  In this section we investigate
systematic problems of the analyzed sample such as boundary effect,
spatial resolution, and smoothing scale.  Section 4 is devoted to the
analysis of volume--limited samples, while the corresponding results
for the apparent--magnitude limited samples are discussed in Appendix
B.  Among other issues we address is the fair sample question,
including a convergence study by comparing with a previous data set.
Finally, Section 5 is devoted to a summary of the results and further
discussion.

\section{SDSS Galaxy and N--body Mock Catalogues }

\subsection{SDSS Sample}

York et al. (2000) provide an overview of the SDSS. Stoughton
et al. (2002) describe the Early Data Release (EDR) and details about
the measurements.  Technical articles providing details of the SDSS
include the description of the photometric camera (Gunn et al. 1998),
photometric analysis (Lupton et al. 2002; Stoughton et al. 2002), the
photometric system (Fukugita et al.  1996; Hogg et al. 2001; Smith et
al. 2002), astrometric calibration (Pier et al. 2002), selection of the
galaxy spectroscopic samples (Strauss et al. 2002; Eisenstein et al.
2001), and spectroscopic tiling (Blanton et al. 2002).

Our analysis of the MFs in the present paper is based on a subset of the
SDSS galaxy redshift data, `Large--scale Structure Sample 12' (Blanton
et al. 2002). This sample includes
galaxies with r--band magnitudes
between $14.5$ and $17.5$ after correction for Galactic reddening using
the maps of Schlegel, Finkbeiner, \& Davis (1998).

A map of the galaxy distribution of the data that we analyze, together with a
typical slice, are shown in Figures \ref{fig:sdssmap_equator},
\ref{fig:sdssmap_survey} and \ref{fig:slicemap}.
The three--dimensional map centered on us in
equatorial coordinate system is shown in the
upper panel of Figure \ref{fig:sdssmap_equator}. The lower panel in
this figure shows the
map projected onto the celestial sphere.
The projected skymap for each Region is plotted in 
the survey coordinate system in Figure \ref{fig:sdssmap_survey}.
Redshift slices of galaxies centered around the equatorial plane
with various redshift limits and thicknesses of planes
are shown in Figure \ref{fig:slicemap}:
$z<0.05$ with thickness of $10h^{-1}$Mpc centered around the
equatorial plane in the upper-left panel;
$z<0.1$ with thickness of $15h^{-1}$Mpc 
in the upper-right panel; 
$z<0.2$ with thickness of $20h^{-1}$Mpc in the lower panel.

The data are mainly located in three regions, which we call Region 1,
2 and 3, respectively.  The properties of each region covering the
range of right ascension $\alpha$ and declination $\delta$ of the
equatorial coordinate system, and the number of galaxies are listed in
Table \ref{tab:region}.  SDSS photometry is taken in driftscan mode
along great circles along arcs of constant $\eta$ in the survey
coordinate system $(\lambda, \eta)$.  The survey coordinate system
$(\lambda,\eta)$ is suitable to describe the survey area, because each
drift scan is perpendicular to a line of constant $\mu$.  The
transformation between celestial and survey coordinates is
\begin{eqnarray}
\label{eq:coordinate}
\cos(\alpha-95^\circ)\cos\delta&=&-\sin\lambda \nonumber \\
\sin(\alpha-95^\circ)\cos\delta&=&\cos\lambda\cos(\eta+32.5^\circ) \\
\sin\delta&=&\cos\lambda\sin(\eta+32.5^\circ) \;.\nonumber 
\end{eqnarray} 
In this paper we do not present the results of our analysis of the
data located in Region 3 (Southern sky) containing the three
$2.5^\circ$ width consecutive stripes separated from each other.  The
reason is that no additional information on the morphology of
three--dimensional structure could be deducted from this set due to
its slice--like geometry.  We also do not use the data in the discrete
area in Region 2 ($22.8^\circ \le \lambda \le 36.1^\circ$, $46.2^\circ
\le \eta \le 51.3^\circ$) containing $3056$ galaxies (Middle panel in
Figure \ref{fig:sdssmap_survey}).

\begin{table*}[h]
\caption{Properties of three main regions of the SDSS `Sample 12',
including the range of each region in the equatorial coordinate system 
($\alpha,\delta$) and the survey coordinate system ($\lambda,\eta$)
(Eq.[\ref{eq:coordinate}]).}
\begin{center}
\begin{tabular}{cccccc}
  \hline\hline
 Name & $\alpha$ & $\delta$ & $\lambda$ & $\eta$ & $N_{\rm gal}$  \\ \hline
 Region 1 & $129.7^\circ\sim250.0^\circ$ & $ -3.74^\circ\sim6.25^\circ$ 
 & $-55.2^\circ\sim65.0$ & $323.7^\circ\sim333.8^\circ$ & $55897$ \\ 
 Region 2 & $112.0^\circ\sim260.6^\circ$ & $23.1^\circ\sim68.8^\circ$ 
 & $-59.8^\circ\sim60.4^\circ$ & $16.2^\circ\sim51.3^\circ$ & $59766$ \\ 
 Region 3 & $308.7^\circ\sim63.3^\circ$ & $-11.3^\circ\sim16.3^\circ $ 
 & $-57.7^\circ\sim55.8^\circ$ & $130.6^\circ\sim159.9^\circ$ & $38624$
 \\ \hline
\end{tabular}
\end{center}
\label{tab:region}
\end{table*}

\begin{figure*}[tph]
\caption{{\it Upper:} 
3D redshift-space map centered on us, and its projection on the
celestial sphere of SDSS galaxy `Sample 12', including the three main
regions listed in Table \ref{tab:region}.  {\it Lower:}
Projected skymap of `Sample 12' galaxy data in equatorial coordinates
($\alpha, \delta$).} 
\label{fig:sdssmap_equator}
\end{figure*}

\begin{figure*}[tph]
\caption{{\it Upper:} 
Projected skymap of `Sample 12' galaxy data in the survey coordinate system
($\lambda, \eta$) in Region 1 ({\it Top}), Region 2 ({\it Middle})
and Region 3 ({\it Bottom}).} 
\label{fig:sdssmap_survey}
\end{figure*}

\begin{figure*}[tph]
\caption{Redshift slices of `Sample 12' galaxy data around the
  equatorial plane. The redshift limits and the thickness of the
  planes are: {\it Upper-left} $z<0.05$, $10h^{-1}$Mpc; {\it
  Upper-right} $z<0.1$, $15h^{-1}$Mpc; {\it Lower} $z<0.2$,
  $20h^{-1}$Mpc.  The size of points has been adjusted. Note that the
  data for the Southern part are sparser than those for the Northern
  part, especially for thick slices (see Figure
  \ref{fig:sdssmap_equator}).}
\label{fig:slicemap}
\end{figure*}

\subsection{Mock samples}

To test for several observational effects on the MFs, including the
shape of the survey volume and the redshift distortion, we construct
mock galaxy samples from a series of P${}^3$M $N$--body simulations
provided by Jing and Suto (1998).  These simulations employ $256^3$
particles in a $(300h^{-1}{\rm Mpc})^3$ periodic comoving box using
Gaussian initial conditions and a Cold Dark Matter (CDM) transfer
function (Bardeen et al. 1986).  We use the $z=0$ snapshot simulation
data (for simplicity we neglect the light--cone effect) in two CDM
cosmological models: Lambda--CDM (LCDM) with
$\Omega_0=0.3,~\lambda_0=0.7,~h=0.7,~\Gamma=0.21$, and $\sigma_8=1$,
and Standard--CDM (SCDM) with
$\Omega_0=1,~\lambda_0=0,~h=0.5,~\Gamma=0.5$, and $\sigma_8=0.6$,
where $h$ denotes the Hubble constant in units of $100$km
s${}^{-1}$Mpc${}^{-1}$, $\Gamma$ is the shape parameter of the
transfer function, and $\sigma_8$ is the r.m.s. density fluctuation
amplitude at $8h^{-1}$Mpc. To simulate the effect of the shape of the
survey volume, we extract wedge samples (12 realizations in total) out
of the full simulation cube so that they have the same sample--shape
and number of particles as each volume--limited sample.  To construct
mock samples that extend beyond the simulation box size, we duplicate
particles using the periodic boundary conditions.  We also consider
the redshift distortion effect by adding the line-of-sight component
of the peculiar--velocity to each particle in the calculation of the
redshift.

\section{The Minkowski Functionals (MFs)}

\subsection{Mathematical Aspects of Minkowski Functionals}

The morphological properties of an $n$--dimensional pattern are
completely described in terms of $n+1$ quantities, which we call MFs.
In the present analysis, we generate isodensity contours from the
three--dimensional density contrast field $\delta$ by taking its
excursion set $F_{\nu}$, i.e., the set of all points where the density
contrast $\delta$ exceeds the threshold level $\nu$.  The four
Minkowski Functionals $V_k(\nu)$ of the excursion set can be measured,
and plotted as functions of the threshold $\nu$.  All MFs can be
interpreted as well--known geometric quantities, namely the
volume fraction $V_0(\nu)$, the total surface area $V_1(\nu)$, the
integral mean curvature $V_2(\nu)$, and the integral Gaussian
curvature, i.e., the Euler characteristic $V_3(\nu)$.

All MFs can be expressed as integrals over the excursion set.  While
the first MF is simply given by the volume integration of a Heaviside
step function $\Theta$ normalized to the total volume $V_{\rm tot}$,
\begin{equation}
V_0(\nu)=\frac{1}{V_{\rm tot}}\int_V d^3x\Theta(\nu-\nu(x))\;,
\end{equation}
the other MFs, $V_k(k=1,2,3)$, are calculated by the surface
integration of the local MFs, $v^{\rm loc}_k$ (Schneider 1978).  The
general expression is
\begin{equation}
\label{eq:local_MFs}
V_k(\nu)=\frac{1}{V_{\rm tot}}\int_{\partial F_{\nu}}
d^2S(\mbox{\boldmath $x$})v^{\rm loc}_k(\nu, \mbox{\boldmath $x$}),
\end{equation}
with the local Minkowski Functionals for $k=1,2,3$ given by
\begin{eqnarray}
v^{\rm loc}_1(\nu,\mbox{\boldmath $x$}) &=& \frac{1}{6}, \\
v^{\rm loc}_2(\nu,\mbox{\boldmath $x$})
&=& \frac{1}{6\pi}\left(\frac{1}{R_1}+\frac{1}{R_2}\right), \\
v^{\rm loc}_3(\nu,\mbox{\boldmath $x$}) &=& \frac{1}{4\pi}\frac{1}{R_1R_2} ,
\end{eqnarray}
where $R_1$ and $R_2$ are the principal radii of curvature of the
isodensity surface.

For a 3--D Gaussian random field, the average MFs per unit volume can be
expressed analytically as follows (Tomita 1990):
\begin{eqnarray}
\label{eq:minkowski_v0}
V_0(\nu) &=&
\frac{1}{2}-\frac{1}{\sqrt{2\pi}}\int^{\nu}_0 \exp{\left(
-\frac{x^2}{2}\right)}dx, \\
\label{eq:minkowski_v1}
V_1(\nu) &=& 
\frac{2}{3}\frac{\lambda}{\sqrt{2\pi}}\exp\left(-\frac{1}{2}\nu^2\right)
\\ 
\label{eq:minkowski_v2}
V_2(\nu) &=& \frac{2}{3}\frac{\lambda^2}{\sqrt{2\pi}}\nu\exp
\left(-\frac{1}{2}\nu^2\right), \\
\label{eq:minkowski_v3}
V_3(\nu) &=& 
\frac{\lambda^3}{\sqrt{2\pi}}(\nu^2-1)\exp\left(-\frac{1}{2}\nu^2\right) ,
\end{eqnarray}
where $\lambda=\sqrt{\sigma_1^2/6\pi\sigma^2}$,
$\sigma\equiv\langle\delta^2\rangle^{1/2}$,
$\sigma_1\equiv\langle|\nabla\delta|^2\rangle^{1/2}$, and $\delta$ is
the density contrast.

In this paper we evaluate the MFs as a function of the threshold level.
We employ two different definitions of this threshold: one is the density
threshold level denoted by $\nu_{\sigma}$, which is given by the density
contrast $\delta$ divided by the r.m.s. density fluctuation $\sigma$
after smoothing, the other is labeled by $\nu_{\rm f}$
parameterizing the volume--fraction $f$ \citep{GMD1986},
\begin{equation}
\label{eq:nuf}
f=\frac{1}{\sqrt{2\pi}}\int^\infty_{\nu_{\rm f}} e^{-x^2/2}dx .
\end{equation}
The meaning of each definition is discussed in the section of the
results for the MFs with volume--limited samples (\S
\ref{sec:result_vollim}).

\subsection{Computational Methods}
\label{subsec:compute_MF}

We compute the MFs for density contrast fields from mock samples and
SDSS galaxies.  We use the cloud--in--cell (CIC) interpolation to assign
survey galaxies and dark matter particles to densities defined on a
$128\times 256\times 64$ grid.  We Fourier transform the density
contrast field, multiply by the Fourier transform of a Gaussian
window with smoothing scale $R_{\rm G}$,
\begin{equation}
W_{\rm G}(r)=\frac{1}{\sqrt{2\pi}R_{\rm G}}
\exp\left(-\frac{r^2}{2R_{\rm G}^2}\right),
\end{equation}
and then transform back to real space.

The accuracy of the estimation of the MFs increases in a specific
volume as we choose smaller values of $R_{\rm G}$, because the
amplitudes of the MFs are roughly proportional to the number of
structures with typical scale $R_{\rm G}$.  On the other hand, in
order to produce reliable results of the MFs, $R_{\rm G}$ should be
bounded by a minimum value for each sample to satisfy the following
criteria (Hoyle et al. 2002):
\begin{enumerate}
\item
$R_{\rm G}$ should be comparable to or larger than the mean separation of
galaxies (using too small values of $R_{\rm G}$ reduces to spherical isodensity contours around each galaxy.)
\item
The r.m.s.~density fluctuation at the smoothing scale of $R_{\rm G}$
should be larger than the Poisson term, which is equal to the inverse
square root of the galaxy number within a Gaussian ball of effective
radius $R_{\rm G}$.
\item
$R_{\rm G}$ should be more than twice the mesh size,
corresponding to the Nyquist resolution frequency.
\end{enumerate}

A useful way to check for systematic errors due to numerical
approximations inherent to each numerical code as well as due to 
the influence of the survey boundary is to compare the results
with another code based on a different computational method.  
There are several interesting routines to estimate the MFs, 
(e.g., Sheth et al. 2003), however, we here use
two well-studied complementary routines to compute the MFs 
of a grid density field.

The first approach transforms the surface integrals in Equations
(\ref{eq:local_MFs}) into volume integrals.  The local Minkowski
Functionals are expressed in terms of invariants formed from the first
and second derivatives of the density contrast field ({\it cf.}
Schmalzing \& Buchert 1997 for an outline of this approach and
Koenderink 1984, and Appendix A for the gory details).  
Hereafter we call this approach {\it Koenderink invariants}.

The other routine is based on Crofton's formula (Crofton 1868) from
integral geometry (Hadwiger 1957). A more detailed description
of the method can be found in Schmalzing \& Buchert 1997. 
This method, which we will refer to as {\it
Crofton's formula}, is best suited for a pattern
described as a set of empty and occupied cells of a cubic grid.  The
calculation of Minkowski Functionals then reduces to counting the
elementary cells (in three dimensions, the points, lines, squares, and
cells of the cubic lattice), that are occupied by the pattern.

For the computation of the fourth MF $V_3$, we also
apply the CONTOUR~3D code (Weinberg 1988), which has been widely used to
calculate the genus $g=-V_3$ (e.g., Gott et al. 1989). 
This code computes the integrated Gaussian
curvature over the surface by considering the dataset to consist of
cubic pixels, so the contour surface consists of faces, edges and
vertices, and where all the curvature is concentrated at the vertices.
The curvature integral is then calculated by summing the angle deficits
(and subtracting the angle excesses) at the vertices.

Although the above three methods are theoretically equivalent, the
numerical results are not always the same.  This is partly due to the
finite cell size of the grid used for our computations. However, the
main difference comes from the effect of the complicated shape of the
survey boundaries.  In the following analysis we test the effect of
excluding part of the volume near the boundary from our analysis, to
find the limited spatial region in which the results from all three methods
are consistent with each other.

\subsection{Log--normal model as a standard of reference}

\begin{figure*}[tph]
\begin{center}
\FigureFile(160mm,160mm){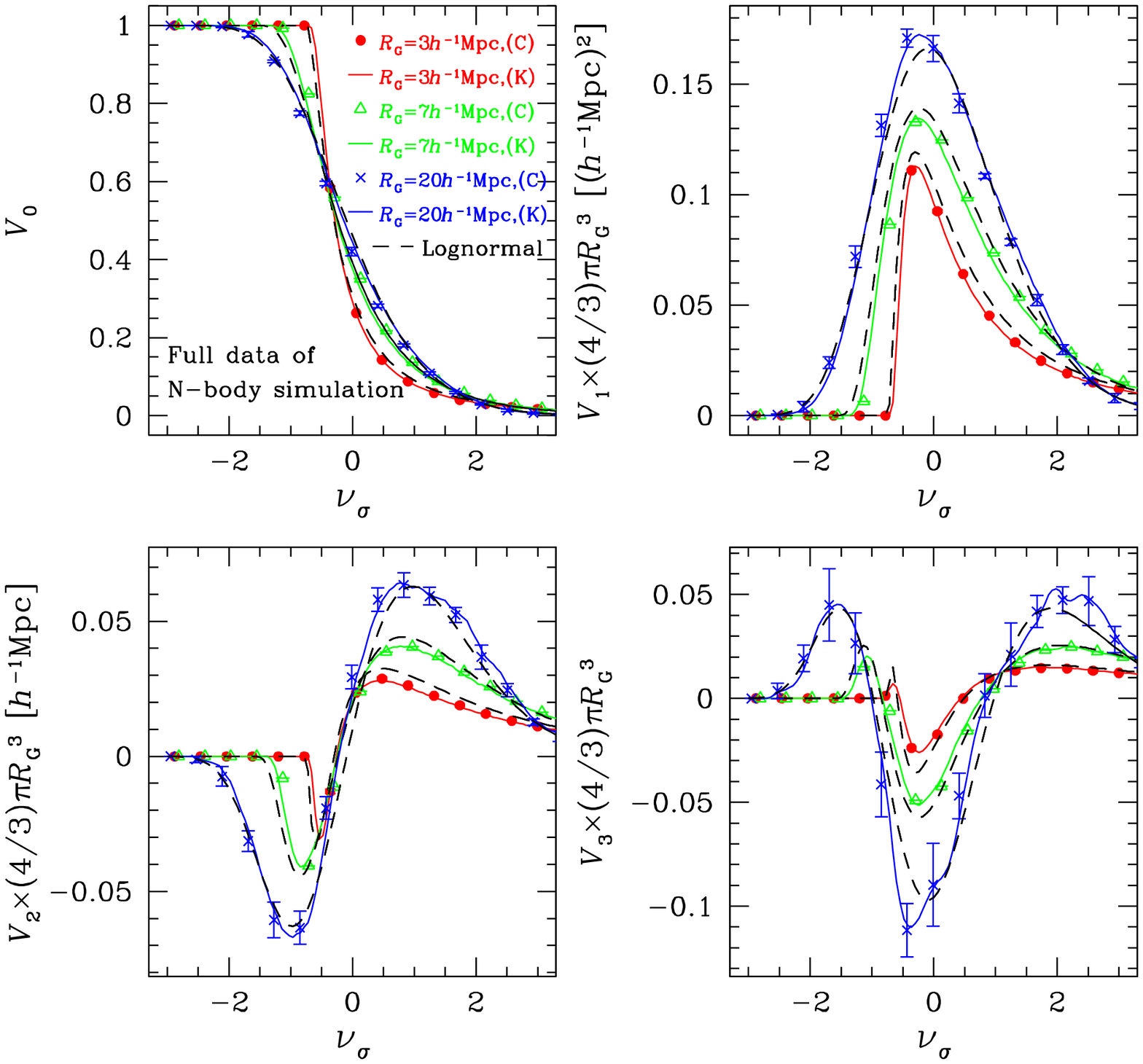}
\end{center}
\caption{Comparison of the MFs for dark matter using all particles
in N-body simulations (($300h^{-1}{\rm Mpc})^3$) run
by Jing \& Suto (1998) with predictions of the log--normal model. 
Here we show estimates of the MFs using two methods:
Crofton's formula labeled by (C) at $R_{\rm G}=$
$3h^{-1}$Mpc (filled circles), $7h^{-1}$Mpc (open triangles), and 
$20h^{-1}$Mpc (crosses), and the Koenderink invariants labeled by (K) at
each smoothing scale (solid lines).
The log--normal model predictions (Eq: (\ref{eq:log-normal}))
are plotted with dashed lines at each smoothing scale. 
All of MFs except for $V_0$ are multiplied by the volume of a Gaussian ball at 
the radius of $R_{\rm G}$, $4/3\pi R_{\rm G}^3$.
Error bars estimated from three realizations of cubic data 
are shown on the results from Crofton's formula.}
\label{fig:full}
\end{figure*}

The genus for the distribution of dark matter in
simulations has been found to be well--approximated by the
log--normal model (Matsubara and Yokoyama 1996; Hikage et al. 2002). 
Therefore, we might expect that this model would provide a reasonable
approximation to all of the MFs of the dark matter distribution.
Matsubara and Yokoyama (1996) derived the
genus expression assuming that the nonlinear density field of dark
matter has a one--to--one correspondence to the primordial Gaussian
field. Generalizing their formula for the genus, the expressions for MFs
that obey the log--normal statistics are obtained by substituting
$\nu_{\rm LN}$ and $\lambda_{\rm LN}$ for $\nu_{\sigma}$ and $\lambda$ in
Equations (\ref{eq:minkowski_v0}) to (\ref{eq:minkowski_v3}):
\begin{eqnarray}
\label{eq:log-normal}
\nu_{\rm LN}(\nu_{\sigma}) &\equiv& 
\frac{\ln[(1+\nu_{\sigma}\sigma)\sqrt{1+\sigma^2}]}
{\sqrt{\ln(1+\sigma^2)}}, \\
\lambda_{\rm LN} &\equiv& \left(\frac{\sigma}{1+\sigma^2\log(1+\sigma^2)}
\right)^{1/2}\lambda .
\end{eqnarray}
Figure \ref{fig:full} shows the comparison between the log--normal
model and the MFs of mock simulations estimated from both the
Koenderink invariants and Crofton's formula methods at $R_{\rm G}= 3,
7, 20h^{-1}$Mpc.  All of MFs except for $V_0$ are multiplied by the
volume of a Gaussian ball at the radius of $R_{\rm G}$, $4/3\pi R_{\rm
G}^3$. The error bar represents the statistical error estimated from
three realizations of the full simulation. We find that all MFs
obtained with the two methods agree very well over a wide range of
smoothing scales.
The log--normal model nicely reproduces all of the four MFs for dark
matter and thus the difference between the log--normal model and the
MFs for dark matter is negligible in the analysis of our galaxy
samples with $10^2\sim10^3$ times smaller volume (see Table
\ref{tab:vollim1}) than that of the full cubic data.

\section{Analysis of Volume--limited Samples}
\label{sec:result_vollim}

\subsection{Construction of Volume--limited Samples}

In an apparent-magnitude limited catalogue of galaxies, the average
number density of galaxies decreases with distance because only
increasingly bright galaxies are included in the sample at larger
distance.  To avoid this systematic change in both density and galaxy
luminosity we construct volume--limited samples of galaxies, with cuts
on both absolute--magnitude and redshift (we do not attempt to correct
for evolution in the galaxy population over these limited ranges of
redshift).  In this section we mainly study the dependencies on galaxy
luminosity and morphological type of the MFs for the volume--limited
samples, and compare with the corresponding mock catalogues.
\begin{figure*}[tph]
\begin{center}
\FigureFile(80mm,80mm){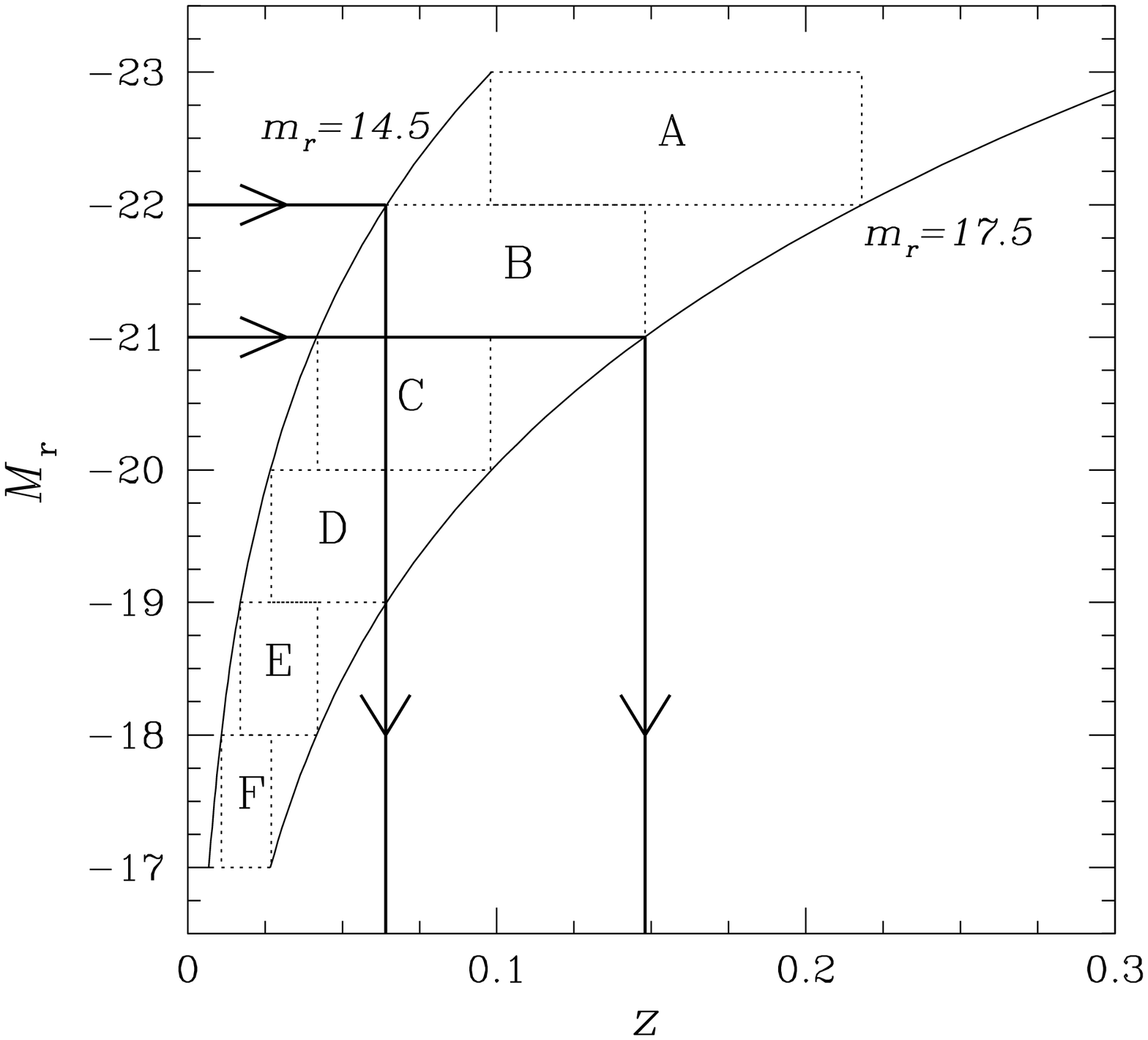}
\end{center}
\caption{Schematic picture to show how we determine the redshift range
of a volume--limited sample with a given absolute--magnitude range.  Six
boxes represent the absolute--magnitude ranges and the redshift ranges
of volume--limited samples listed in Table \ref{tab:vollim1}.  Solid
lines indicate the relation between $M_{\rm r}$ and $z$ (Eq. (\ref{eq:Mrtoz}))
at apparent magnitudes $m_{\rm r}=14.5$ and $17.5$.}  \label{fig:Mrtoz}
\end{figure*}
We construct six volume--limited samples in Region 1 and Region 2
covering ranges of absolute magnitude of width $\Delta M_{\rm r}=1$
from $-23$ to $-17$, where the absolute magnitude is
\begin{equation}
\label{eq:Mrtoz}
M_{\rm r}=m_{\rm r}-5\log[r(1+z)]+K(z)\;,
\end{equation}
and where $K(z)$ is the K-correction for the bandpass shift 
due to the redshift.
We apply an approximate K--correction factor $K(z) = 0.9z$
valid for the typical galaxy color of $g - r=0.65$ (Fukugita et al. 1995).
The maximum (minimum) redshift is
determined by the redshift at which the faintest (brightest)
galaxies in each sample lies in the apparent--magnitude range
$14.5<m_{\rm r}<17.5$ (see Figure \ref{fig:Mrtoz}).  
The bright cutoff in this apparent magnitude range approximates the
three arcsecond aperture magnitude limit in the spectroscopic target
selection (see Strauss et al. 2002 for details).
We compute
the comoving distance from the observed redshift $z$ of each galaxy after
correction for the Local Group motion,
\begin{equation}
\label{eq:rz}
\hspace*{-1cm}
d_c(z)=\int_0^{z}\frac{dz^\prime}
{\sqrt{\Omega_0(1+z^\prime)^3+(1-\Omega_0-\lambda_0)(1+z^\prime)^2
+\lambda_0}} \;,
\end{equation}
where $\Omega_0$ is the matter density parameter and $\lambda_0$ is the
dimensionless cosmological constant.
Table \ref{tab:vollim1} shows the redshift range and the total volume
of each volume--limited sample.
\begin{table*}[h]
\caption{Redshift range and survey volume in Regions 1 and 2 
for each volume limited--sample with different magnitude ranges.}
\begin{center}
\begin{tabular}{ccccc}
\hline\hline
& & & \multicolumn{2}{c}{total volume [$(h^{-1}{\rm Mpc})^3$]} \\ 
\raisebox{1.5ex}[0pt]{sample name} & 
\raisebox{1.5ex}[0pt]{magnitude range} & \raisebox{1.5ex}[0pt]{redshift range} 
& Region 1 & Region 2  \\ \hline
A & $-23<M_{\rm r}<-22$ & $0.098<z<0.218$ 
& $1.62\times 10^7$ & $1.86\times 10^7$  \\
B & $-22<M_{\rm r}<-21$ & $0.064<z<0.148$ 
& $5.40\times 10^6$ & $6.17\times 10^6$  \\
C & $-21<M_{\rm r}<-20$ & $0.042<z<0.098$ 
& $1.66\times 10^6$ & $1.90\times 10^6$  \\
D & $-20<M_{\rm r}<-19$ & $0.027<z<0.064$ 
& $4.81\times 10^5$ & $5.50\times 10^5$  \\
E & $-19<M_{\rm r}<-18$ & $0.017<z<0.042$ 
& $1.33\times 10^5$ & $1.52\times 10^5$  \\
F & $-18<M_{\rm r}<-17$ & $0.011<z<0.027$ 
& $3.57\times 10^4$ & $4.08\times 10^4$ \\ \hline
\end{tabular}                                                               
\end{center}
\label{tab:vollim1}
\end{table*}

\begin{table*}[h]
\caption{Properties of volume--limited samples with different 
absolute-magnitude range, including the number of galaxies 
and the mean separation for All, Early--type and
Late--type galaxies, respectively.}
\begin{center}
\begin{tabular}{cccccc}
\hline\hline
& & \multicolumn{2}{c}{number of galaxies}
& \multicolumn{2}{c}{mean separation [$h^{-1}{\rm Mpc}$]} \\ 
\raisebox{1.5ex}[0pt]{sample name} & \raisebox{1.5ex}[0pt]{galaxy type}
& Region 1 & Region 2  & Region 1 & Region 2  \\ \hline
A &   All & $ 853$ & $ 907$ & $26.7$ & $27.4$ \\
  & Early & $ 592$ & $ 645$ & $30.1$ & $30.7$ \\ 
  &  Late & $ 151$ & $ 131$ & $47.5$ & $52.2$ \\
B &   All & $6727$ & $6286$ & $ 9.3$ & $ 9.9$ \\
  & Early & $3570$ & $3221$ & $11.5$ & $12.4$ \\
  &  Late & $1938$ & $1699$ & $14.1$ & $15.4$ \\
C &   All & $9783$ & $7750$ & $ 5.5$ & $ 6.3$ \\
  & Early & $4049$ & $3178$ & $ 7.4$ & $ 8.4$ \\
  &  Late & $3849$ & $3276$ & $ 7.6$ & $ 8.3$ \\  
D &   All & $3677$ & $4844$ & $ 5.1$ & $ 4.8$ \\
  & Early & $ 987$ & $1256$ & $ 7.9$ & $ 7.6$ \\
  &  Late & $2099$ & $2644$ & $ 6.1$ & $ 5.9$ \\  
E &   All & $1648$ & $1671$ & $ 4.3$ & $ 4.5$ \\
  & Early & $ 245$ & $ 253$ & $ 8.2$ & $ 8.4$ \\
  &  Late & $1071$ & $1073$ & $ 5.0$ & $ 5.2$ \\  
F &   All & $ 686$ & $ 416$ & $ 3.7$ & $ 4.6$ \\
  & Early & $  63$ & $  37$ & $ 8.3$ & $10.3$ \\
  &  Late & $ 422$ & $ 258$ & $ 4.4$ & $ 5.4$ \\ \hline
\end{tabular}                                                               
\end{center}
\label{tab:vollim2}
\end{table*}
\begin{table*}[h]
\caption{Properties of density fields for each volume--limited sample
including the smoothing scale $R_{\rm G}$, the absolute--magnitude
range, the mesh size of the simulation box, the resolution number
$N_{\rm res}$ (Eq. (\ref{eq:nres})), and the r.m.s. density fluctuation
amplitude $\sigma$ of the SDSS galaxy number density field and the
averaged $\sigma$ with one--sigma error of the particle number density
field in the mock samples.}
\begin{center}
\begin{tabular}{ccccccccc}
\hline\hline
& sample & & \multicolumn{2}{c}{$N_{\rm res}$}
& \multicolumn{2}{c}{$\sigma$ of SDSS galaxies} 
& \multicolumn{2}{c}{$\sigma$ of mock samples} \\ 
\raisebox{1.5ex}[0pt]{$R_{\rm G}$} & name
& \raisebox{1.5ex}[0pt]{mesh size} 
& Region 1 & Region 2 & Region 1 & Region 2 & Region 1 & Region 2 \\ \hline
$ 3h^{-1}$ Mpc & D & $1.7$ $h^{-1}$Mpc & $1131$ & $1294$ & $1.46$ 
& $1.37$ & $1.38\pm0.06$ & $1.38\pm0.13$ \\
               & E & $1.1$ $h^{-1}$Mpc & $ 312$ & $ 357$ & $1.31$ 
& $1.26$ & $1.28\pm0.13$ & $1.38\pm0.16$ \\
               & F & $0.7h^{-1}$ Mpc & $  83$ & $    95$ & $1.11$ 
& $1.23$ & $1.47\pm0.54$ & $1.68\pm0.58$ \\
$ 5h^{-1}$ Mpc & C & $2.6$ $h^{-1}$Mpc & $ 843$ & $ 965$ & $1.13$ 
& $1.10$ & $0.94\pm0.03$ & $1.00\pm0.07$ \\
               & D & $1.7$ $h^{-1}$Mpc & $ 244$ & $ 279$ & $1.02$ 
& $0.96$ & $0.94\pm0.04$ & $0.95\pm0.09$ \\
               & E & $1.1$ $h^{-1}$Mpc & $  67$ & $  77$ & $0.91$ 
& $0.88$ & $0.89\pm0.10$ & $0.94\pm0.10$ \\
$ 7h^{-1}$ Mpc & C & $2.6$ $h^{-1}$Mpc & $ 307$ & $ 351$ & $0.93$ 
& $0.86$ & $0.72\pm0.03$ & $0.77\pm0.05$ \\
               & D & $1.7$ $h^{-1}$Mpc & $  89$ & $ 101$ & $0.77$ 
& $0.73$ & $0.70\pm0.04$ & $0.73\pm0.07$ \\
               & E & $1.1$ $h^{-1}$Mpc & $  24$ & $  28$ & $0.71$ 
& $0.69$ & $0.70\pm0.10$ & $0.71\pm0.07$ \\
$10h^{-1}$ Mpc & B & $3.8$ $h^{-1}$Mpc & $ 342$ & $ 391$ & $0.69$ 
& $0.77$ & $0.58\pm0.02$ & $0.62\pm0.03$ \\
               & C & $2.6$ $h^{-1}$Mpc & $ 105$ & $ 120$ & $0.74$ 
& $0.64$ & $0.53\pm0.02$ & $0.58\pm0.04$ \\               
               & D & $1.7$ $h^{-1}$Mpc & $  30$ & $  34$ & $0.56$ 
& $0.53$ & $0.51\pm0.04$ & $0.54\pm0.06$
\\ \hline 
\end{tabular}                                                               
\end{center}
\label{tab:vollim3}
\end{table*}

These galaxies are further classified into two morphological types, {\it
Early--type} and {\it Late--type}, which basically correspond to E/S0 and
Sp/Irr, respectively.  \citet{Shimasaku2001} found a tight correlation
between the morphology and the inverse concentration index $c_{\rm
in}$, defined as the ratio of the half--light Petrosian radius to the 90
\%-light Petrosian radius (Stoughton et al. 2002).  
We adopt the threshold values of $c_{\rm
in}=0.35$ for $m_{\rm r}<16.0$, $0.359$ for $16.0< m_{\rm r} <16.5$, and
$0.372$ for $16.5< m_{\rm r} <17.0$. Since the discrimination of
morphology is very difficult for $m_{\rm r}>17.0$, we count such faint
galaxies neither as Early--type nor Late--type (this is why the number
of galaxies for early and late types is different from that of all
galaxies in Table \ref{tab:vollim2}).  Table \ref{tab:vollim2} shows the
properties of each volume--limited sample including the numbers and the
mean separations for Early--type, Late--type and All (including the
faint galaxies with $17.0<m_{\rm r}<17.5$) in each region.

Following the criteria for the smoothing scale $R_{\rm G}$ discussed above,
we choose the three appropriate volume--limited samples and  analyze them with
smoothing lengths $R_{\rm G}=3$,
$5$, $7$ and $10h^{-1}$Mpc, respectively, listed in Table \ref{tab:vollim3}.
The amplitude of the 
the MFs for each smoothed density field varies with the effective
number of resolution elements in the sample
$N_{\rm res}$
\citep{VPGHG1994}:
\begin{equation}
\label{eq:nres}
N_{\rm res}=V_{\rm tot}/(2\pi)^{3/2}R_{\rm G}^3\;,
\end{equation}
where $V_{\rm tot}$ is the total survey volume in which we compute the
MFs.  The estimator $N_{\rm res}$ indicates the number of structures at
a typical scale $R_{\rm G}$ in the smoothed field and is roughly
proportional to the amplitude of the MFs (the sampling noise of the
MFs is roughly proportional to the inverse square root of $N_{\rm res}$). We
also list the value of $N_{\rm res}$ for each smoothed density field in
Table \ref{tab:vollim3}.

We calculate the r.m.s. fluctuation amplitude $\sigma$ for each smoothed
density field.  Table \ref{tab:vollim3} shows that $\sigma$ of the SDSS
galaxies agrees well with that of the dark matter samples within the
standard deviation for all smoothing scales and samples with different
magnitude ranges. This implies that the current data of SDSS galaxies
may already approach a fair sample with respect to two--point measures
on scales $\le 10h^{-1}$Mpc.
We also find that luminous
galaxies exhibit stronger clustering (larger $\sigma$).

\begin{figure*}[tph]
\caption{{\it top:} Illustration of a volume--limited sample C. 
  Upper figures are redshift slice diagrams of
  Early--type (red), Late--type (blue), and Faint ($17.0<m_{\rm
    r}<17.5$) galaxies (black) for Region 1 ({\it left}) and Region 2
  ({\it right}).  Lower panels show redshift-space maps of the
  smoothed density field of the volume--limited sample for All,
  Early--type and Late--type galaxies at $R_{\rm G}=7h^{-1}$Mpc.}
\label{fig:pointset}
\end{figure*}

Figure \ref{fig:pointset} shows projected point distributions and
contour plots of the density fields of the volume--limited samples
smoothed at $R_{\rm G}=7h^{-1}$Mpc for Early--type,
Late--type, and All galaxies, respectively
(see discussion in Subsection \ref{subsec:morphology}).

\begin{figure*}[tph]
\begin{center}
\FigureFile(160mm,160mm){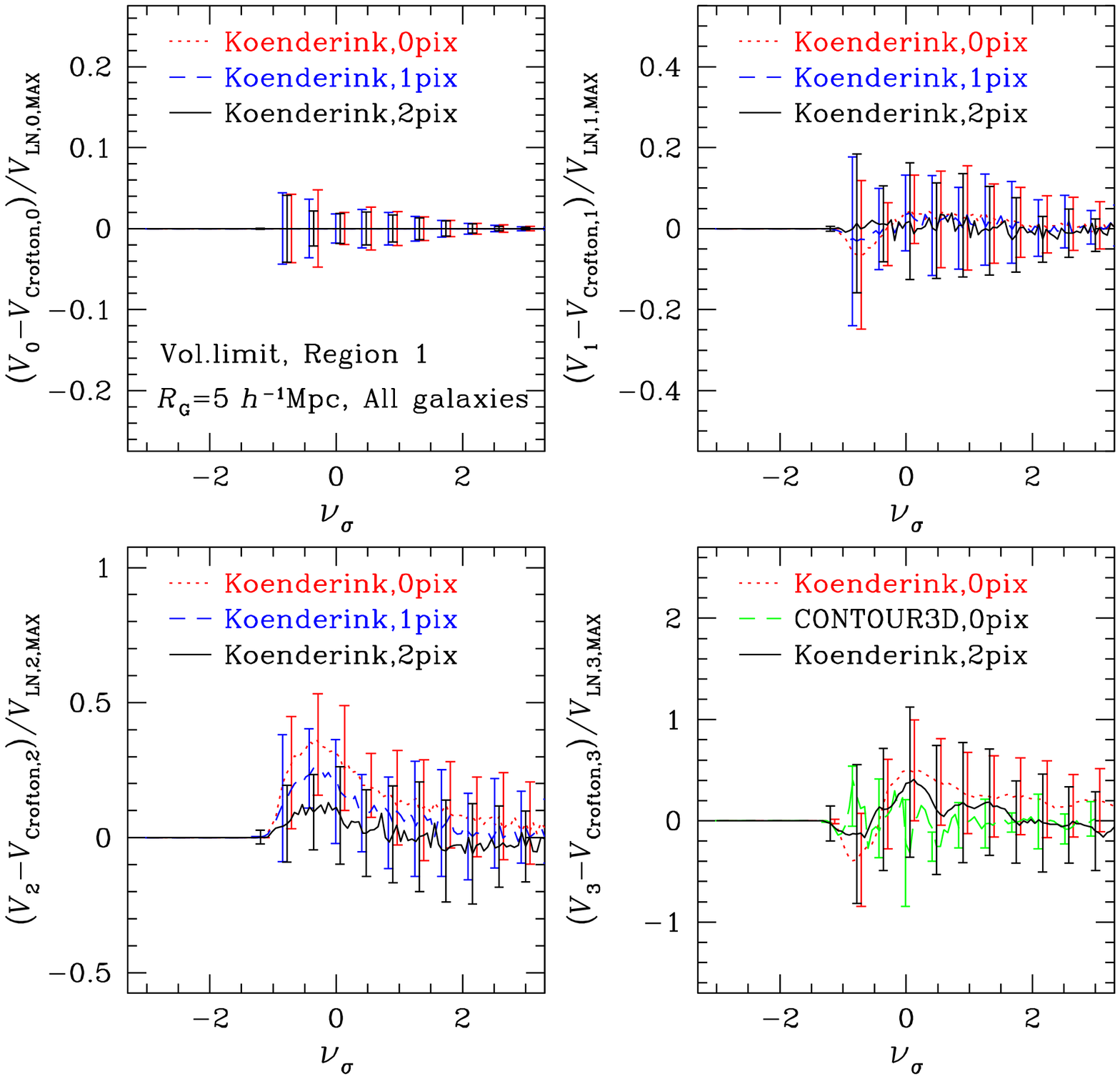}
\end{center}
\caption{Comparison of the two methods used to compute the MFs of
  mock volume--limited samples of Region 1. The plotted lines show the
  difference between the MF estimated using the Koenderink invariants
  method and the Crofton's formula method, normalized by the MF
  amplitude predicted by the log--normal model. Error bars indicate
  the standard deviation estimated from the mock samples.  The region
  near the survey boundary that is excluded from the analysis varies
  from from zero (dotted lines), to one grid cell (dashed lines) or
  two grid cells (solid lines) from the survey edge. For the fourth
  MF, $V_3(\nu_{\sigma})$, the difference between results from the CONTOUR~3D
  code (Weinberg 1998) and Crofton's formula is plotted with dashed
  lines.  The smoothing scale is $R_{\rm G}=5h^{-1}$Mpc, and the
  absolute--magnitude range is $-20<M_{\rm r}<-19$.}
\label{fig:volmagmethod}
\end{figure*}

Differences between estimates of the MFs using Crofton's formula and
the Koenderink invariants methods is shown in Figure
\ref{fig:volmagmethod}. The methods yield consistent results except
for $V_2$. In that case the difference is minimized to an acceptable
level by restricting the region of analysis to a minimum of two grid
cells from the boundary.

Here, two comments are in order: first, $V_2$ is, according to many
previous studies, the most sensitive of the four MFs to the morphology
of large--scale structure. 
When the survey of the north galactic cap is completed, $V_2$ will be
accurately measured. Second, we found that the agreement between both
methods is better for all MFs at smaller smoothing lengths.  For
$V_3(\nu)$, which is the same as the genus except for the sign, we
also plot the difference between the Crofton's formula results and the
genus calculated by CONTOUR~3D (Weinberg 1998).  The agreement between
the two methods is also satisfactory.  Therefore, we show only the
estimation of the MFs based on Crofton's formula for all MFs in what
follows. In order to effectively enlarge the analyzed sample region and so
improve on the significance of the results, we averaged the values
for the MFs for Region 1 and Region 2. In doing so, we also take into
account cosmic variance, to which we devote a separate study in the
next subsection.

\subsection{Cosmic Variance}

\begin{figure*}[tph]
\begin{center}
\FigureFile(100mm,100mm){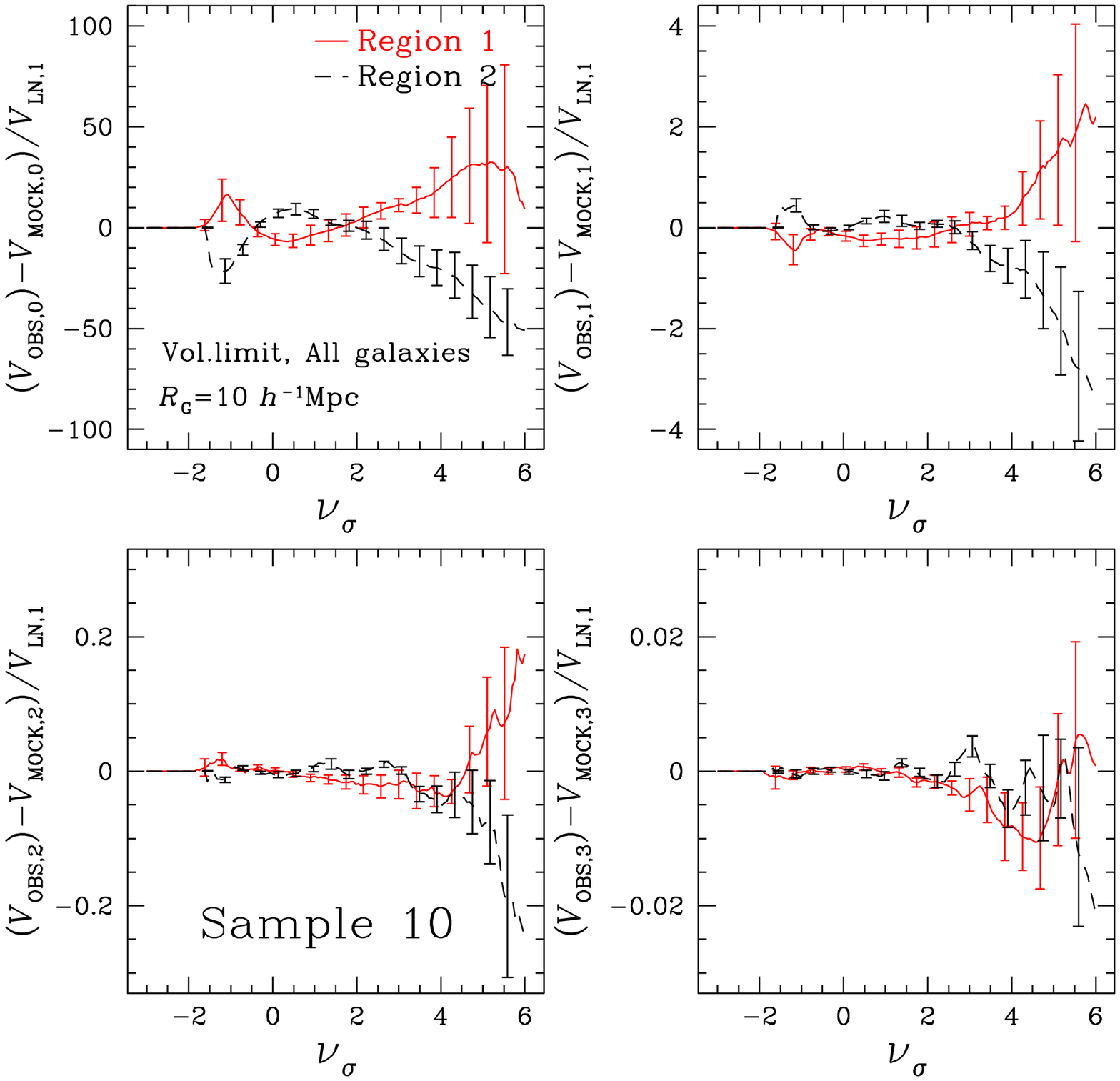}
\FigureFile(100mm,100mm){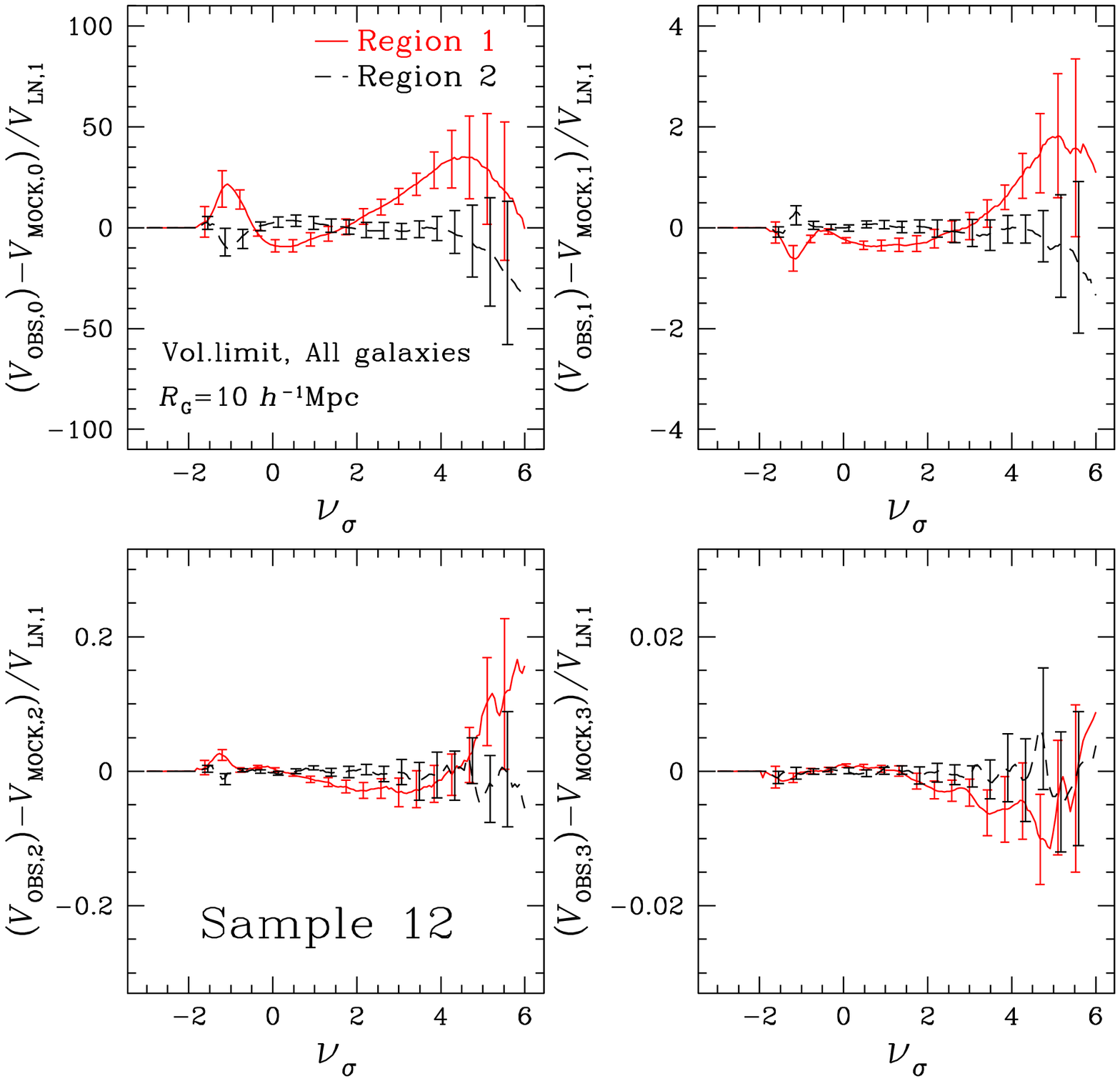}
\end{center}
\caption{Comparison of the MFs for the volume--limited samples of SDSS
  galaxies, Region 1 (solid lines) and Region 2 (dashed lines).
  We plot the difference between the MFs of the observations and mock
  surveys of each region, normalized by the amplitude of
  each MF predicted from the log--normal model for `Sample 10' (upper
  panel) and `Sample 12' (lower panel).  Error bars indicate the
  standard deviation estimated from the mock results.  The smoothing
  scale is $R_{\rm G}=10h^{-1}$Mpc, and the absolute--magnitude range
  is $-20<M_{\rm r}<-19$.}
\label{fig:volregion}
\end{figure*}

To examine whether
Region 1 and Region 2 are fair samples for the purpose of measuring
the MFs,
we plot the difference between the observed MFs, $V_{{\rm obs}, k}$,
and MFs from mock samples $V_{{\rm mock},k}$ for Region 1 and Region 2
simultaneously in Figure \ref{fig:volregion}.  We normalize the
difference by the $V_{{\rm LN}, 1}(\nu)$, which is the second MF
predicted by the log--normal model (Eq. (\ref{eq:log-normal})) to remove
the exponential damping nature of the MFs for large thresholds.  The
error bars on the line indicate the statistical error estimated from 12
realizations of the mock samples for each region.

We also compare the results for `Sample 12' (lower panel in Figure
\ref{fig:volregion}) with the results of the previous `Sample 10' (upper
panel) which contains $\sim 30\%$ fewer galaxies and has $\sim 20\%$
smaller survey volume than `Sample 12'.  We find that all of
the four MFs marginally agree within the statistical error between
Region 1 and Region 2 for `Sample 12', while significant morphological
fluctuations are seen in the MFs for Regions 1 and 2 of `Sample 10'.

This ``convergence property'' of our results by moving from `Sample 10'
to `Sample 12' supports our methods of analysis and indicates that we
obtained statistically reliable results.
In the case of `Sample 12' the variation of the MFs between Regions 1
and 2 is consistent with the uncertainties due to cosmic variance
estimated using the mock samples. However, `Sample 10'
still displays significant morphological discrepancies.  As we learned
from previous studies of the PSCz catalogue (Kerscher et al. 2001)
compared with the IRAS 1.2 Jy catalogue (Kerscher et al. 1998; 1997),
morphological fluctuations on these spatial scales can be
large, and the `fair sample scale' may well be larger than
$300h^{-1}$Mpc. Our previous studies mentioned above also showed that
the sample--to--sample variation in mock catalogues critically depends
on the size of the simulation box; simulations with box--size of the
mock catalogues we employed in the present investigation are not
expected to be able to reproduce the large morphological fluctuations
that could be present on the scales considered. In any case, this is one
of the critical issues that can be answered more reliably with the
future samples of SDSS galaxies.

\subsection{Luminosity Segregation}

\begin{figure*}[tph]
\begin{center}
\FigureFile(95mm,95mm){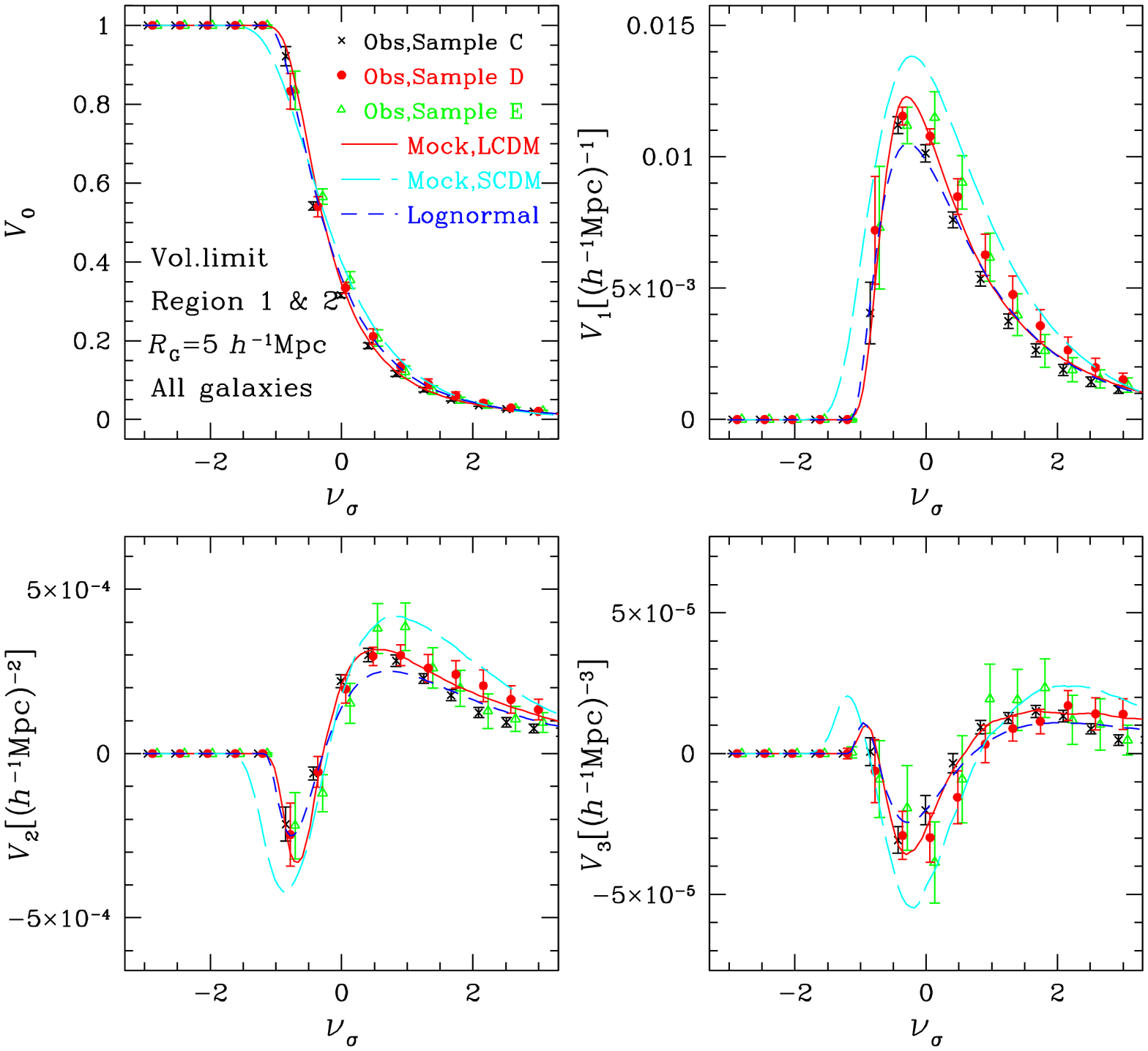}
\FigureFile(95mm,95mm){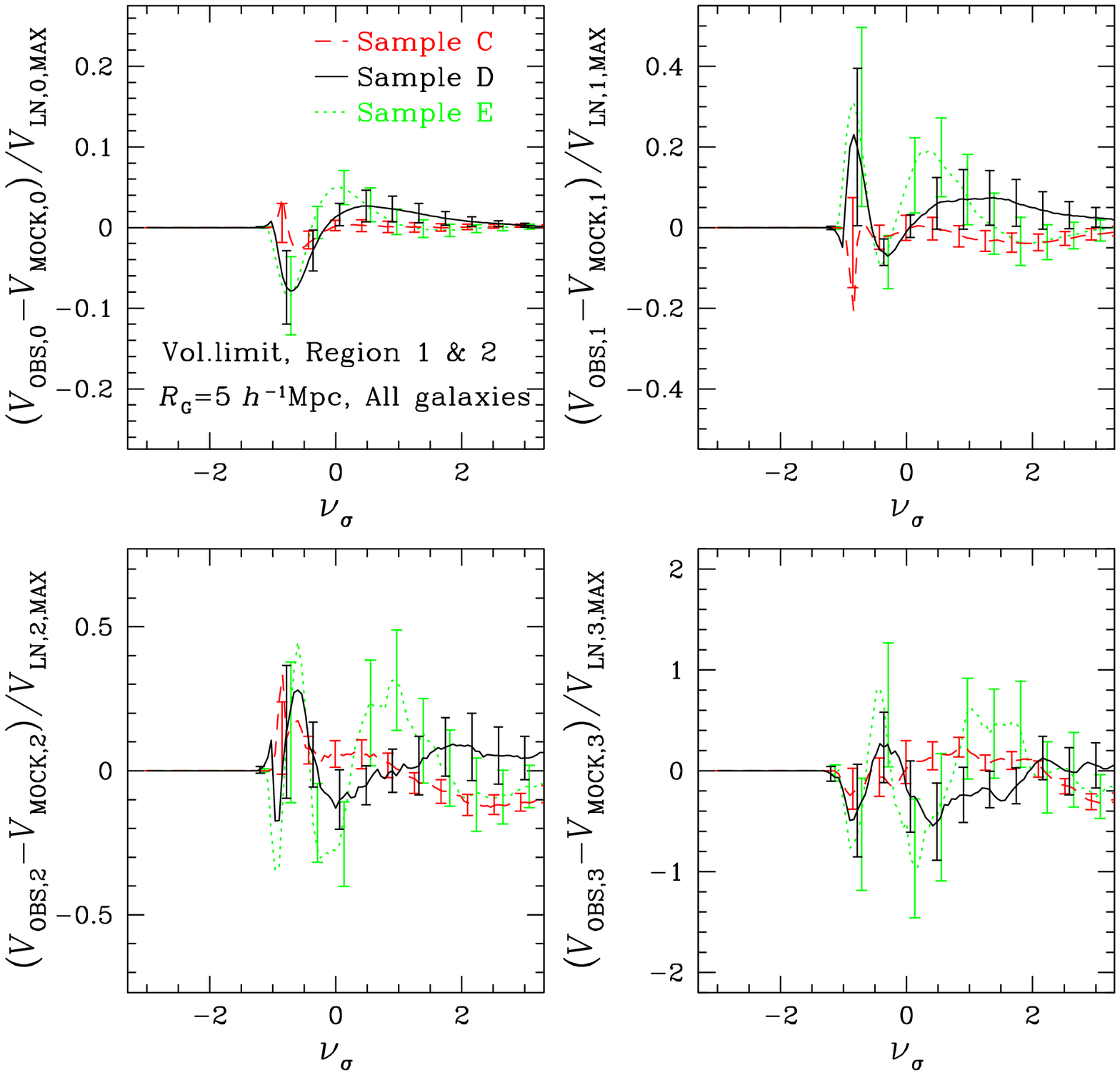}
\end{center}
\caption{Comparison of the MFs of volume--limited samples with
different absolute--magnitude ranges (open triangles, filled circles
and crosses in decreasing orders of $M_{\rm r}$). 
In the right set of panels, the 
lines represent the difference between the observed MFs and results
for the  mock samples from simulations, normalized by the amplitude of each MF
predicted from the log--normal model. 
Sampling errors estimated from mock samples are plotted with the
observational results.  Averaged MFs of the mock samples
are plotted for LCDM (solid lines) and SCDM (long dashed lines).
Log--normal model predictions are also plotted with short dashed lines.
Plotted lines are the average between Region 1 and 2  
at $R_{\rm G}=5h^{-1}$Mpc.}
\label{fig:volmag_mix}
\end{figure*}

In Figure \ref{fig:volmag_mix} we plot the four MFs for the
volume--limited samples at $R_{\rm G}=5h^{-1}$Mpc with three different
absolute--magnitude ranges, $-21<M_{\rm r}<-20$ (Sample C), 
$-20<M_{\rm r}<-19$ (Sample D), and 
$-19<M_{\rm r}<-18$ (Sample E) for Region 1 and 2 together.  
The properties of each volume--limited sample
are listed in Table \ref{tab:vollim3}. The error bars on each observed
MF represent the standard deviation estimated from the mock samples.
For comparison we plot the averaged MFs for the mock samples with
intermediate absolute--magnitude range, that is, Sample D
for both LCDM and SCDM models. The analytical prediction of the
log--normal model (Eq. (\ref{eq:log-normal})) is also plotted with
$\sigma$ and $\sigma_1$ calculated from the fitting formula of the power
spectrum of the LCDM model by Peacock \& Dodds (1996).

Figure \ref{fig:volmag_mix} addresses three issues: first, the
comparison between the log--normal model and the simulated MFs, which
corresponds to the MFs for dark matter, second, the comparison 
between the simulated MFs and the
observed MFs, and third, the comparison among the observed MFs
of galaxy subsamples of different luminosity.  First we find that 
the log--normal predictions agree well with the MFs for dark matter,
even though the shape of the volume is wedge--like. 
This means that the observational effects on the MFs, such as the
survey shape and the redshift distortion, are negligible, at least in
the samples that we analyze. These results are consistent with previous
results on the genus analysis for the SDSS Early
Data Release \citep{Hoyle2002,H2002}.

Next, we find that the observations agree better with the mock results
of the LCDM model than with those for the SCDM model, for all of the
MFs at all smoothing scales considered.  Assuming a scale--invariant
linear bias between dark matter and galaxies, the agreement is
consistent with the $\Lambda$--dominated spatially--flat model with
random--Gaussian initial conditions.

Finally, the luminosity dependence of the MFs is found to be small
compared with the standard deviation.  We plot the difference between the
MFs of the observation and the mock samples (subtraction of the
mock results from the observed MFs at each $\nu$) in the right panels of
Figure \ref{fig:volmag_mix}. We find that there is no clear difference between
the MFs for galaxy samples with different luminosities.  Our
results suggest that the sensitivity of the nonlinearity of the biasing
to the galaxy luminosity is small, in other words, the biasing effect of
galaxy luminosity is negligible, at least on the volume--scale of
`Sample 12'.

\begin{figure*}[tph]
\begin{center}
\FigureFile(160mm,160mm){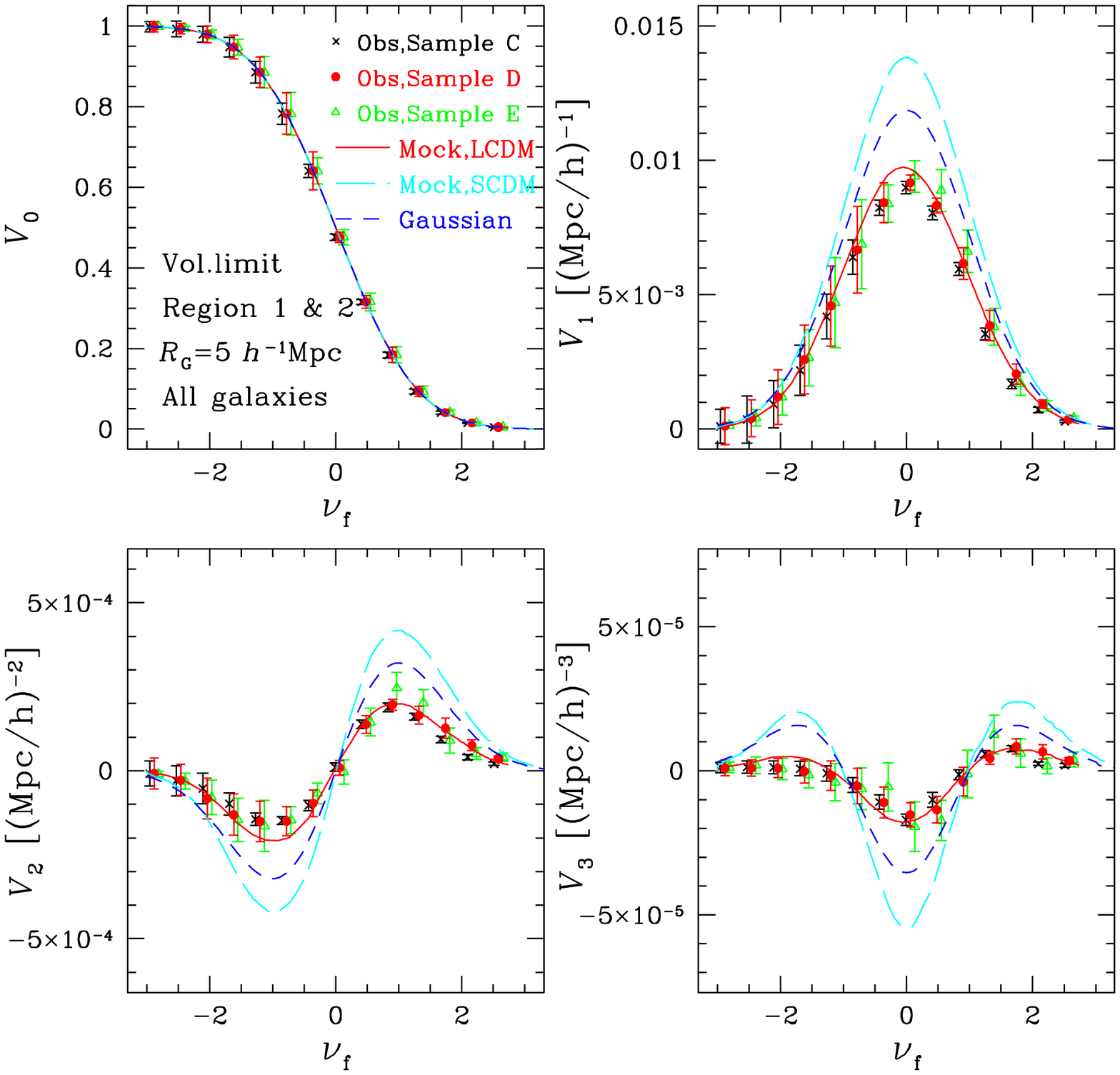}
\end{center}
\caption{MFs as a function of $\nu_{\rm f}$ for $R_{\rm G}=5h^{-1}$Mpc 
in Region 1 and Region 2 together.
(Compare Figure \ref{fig:volmag_mix} for the MFs as a function of 
$\nu_{\sigma}$). }
 \label{fig:volmagvfr}
\end{figure*}

Figure \ref{fig:volmagvfr} shows the MFs as a function of $\nu_{\rm f}$ defined
from the volume fraction (Eq. [\ref{eq:nuf}]).  Instead of the
log--normal model prediction, the analytical predictions in
random--Gaussian statistics are plotted. All of our results favor the
LCDM model with random--Gaussian initial conditions.

\subsection{Morphological Segregation}
\label{subsec:morphology}

\begin{figure*}[tph]
\begin{center}
\FigureFile(95mm,95mm){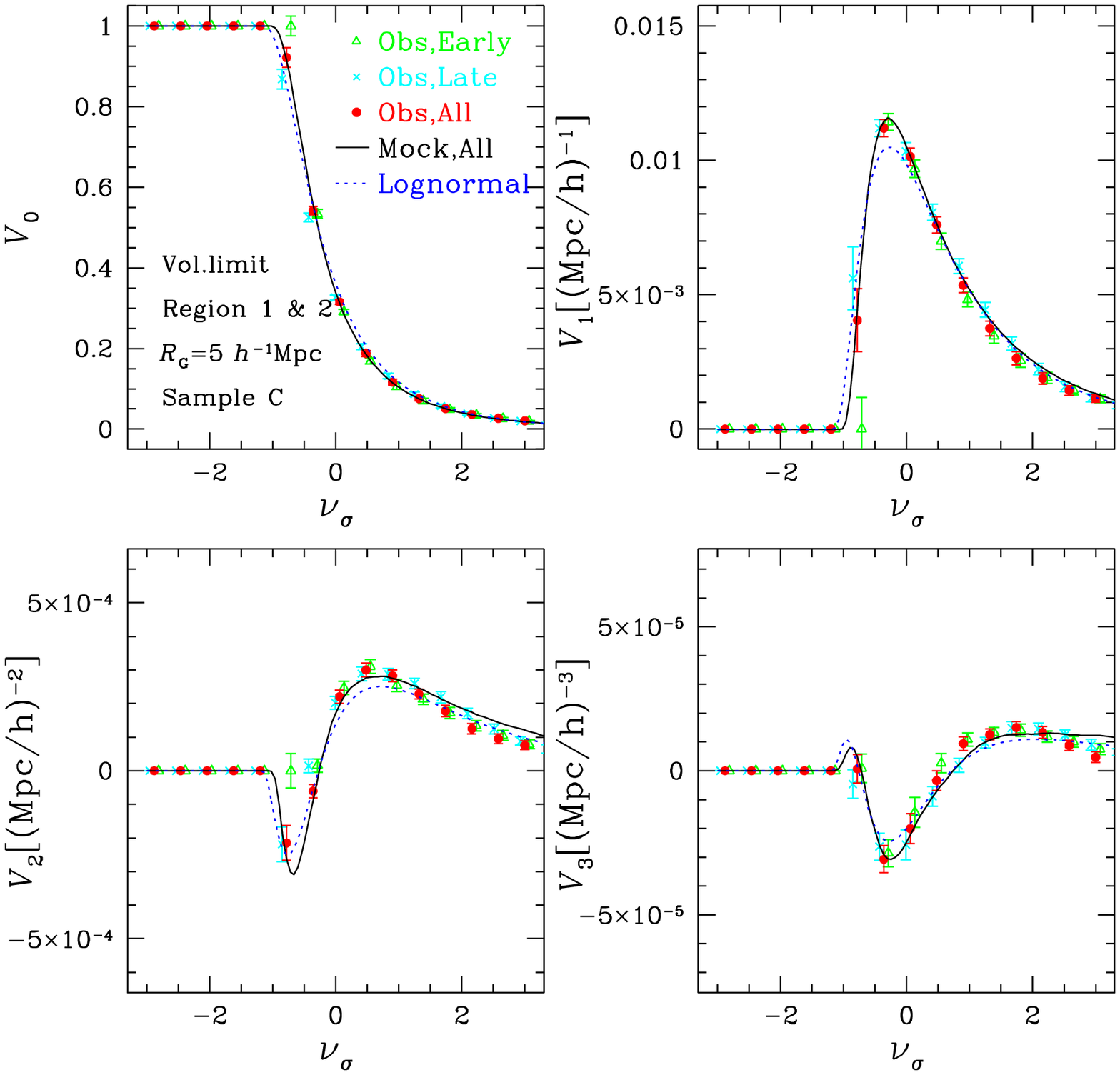}
\FigureFile(95mm,95mm){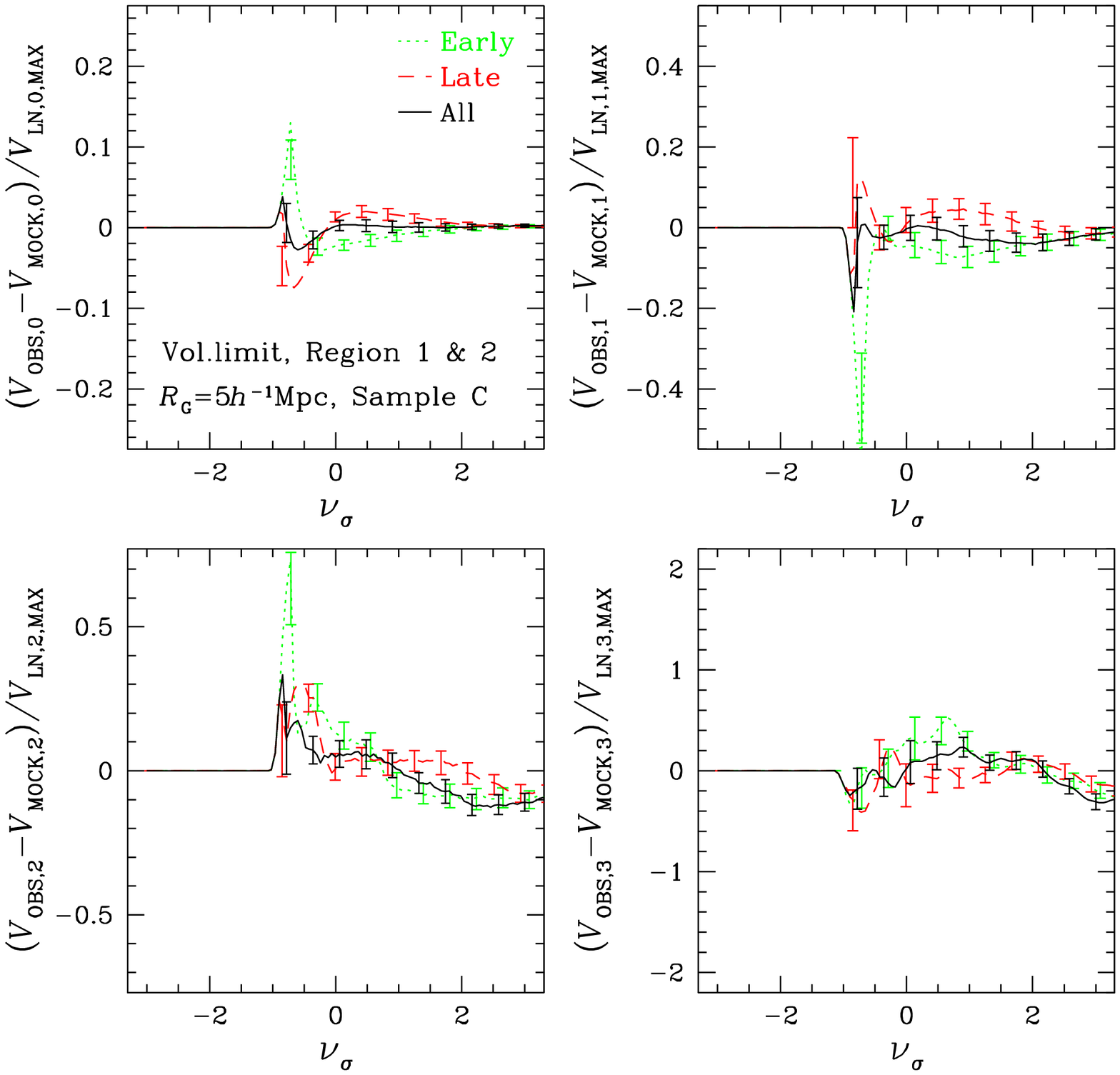}
\caption{Comparison of the MFs of a volume--limited sample C of
Early--type galaxies (open triangles), Late--type galaxies (crosses) and
All galaxies (filled circles). Plotted lines are the average between 
Region 1 and Region 2 smoothed at $R_{\rm G}=5h^{-1}$Mpc. 
The lower panel displays the difference
between the observed and mock sample MFs, normalized by
the amplitude of each MF predicted from the log--normal model.  
Sampling errors estimated from mock samples are plotted with
the observational results.  Averaged MFs of the mock
samples for the distribution of all galaxies based on LCDM are plotted
with solid lines.  Log--normal model predictions are also plotted with
short dashed lines.  \label{fig:voltype} }
\end{center}
\end{figure*}

In this subsection we perform a quantitative analysis of the
dependencies of the MFs on morphological type of galaxies ({\em
cf.}~Figure \ref{fig:pointset}). Similar to the analysis of luminosity
segregation in the previous subsection, we here compare the MFs for
a volume--limited sample, Sample C, smoothed at
$5h^{-1}$Mpc for different morphological types: Early--type,
Late--type, and All galaxies shown in Figure \ref{fig:voltype}. For
comparison we plot the MFs for dark matter based on the LCDM and SCDM
models, as well as the log--normal predictions.

In the lower panel of Figure \ref{fig:voltype}, which shows the
difference between the observed MFs and simulated MFs (the same as the
lower panel of Figure \ref{fig:volmag_mix}), we find that the
variations of the MFs for different galaxy type is small enough to
conclude that morphological biasing of the MFs is negligible in `Sample 12'.

\section{Summary and Discussion}

We present results of a detailed morphometric analysis of a sample of
SDSS galaxy data containing $154844$ galaxies (`Sample 12') employing
the complete set of Minkowski Functionals (MFs). To test for
systematic errors in calculating the MFs of isodensity surfaces, we
employ two complementary computational methods for the MFs: Crofton's
formula and Koenderink invariants. We find that these methods produce
results that agree to within statistical errors, despite the
complicated shape of the survey region. A possible exception to this
agreement occurs for the most sensitive MF, $V_2$, but even in this
case the dependence on method becomes small when the computed region
is limited to several pixels inward from the survey edge.  By
constructing volume--limited samples, in which the distribution of
galaxies in predefined magnitude ranges is homogeneous, we conduct a
detailed analysis of the observed MFs for galaxies within different
magnitude ranges and for galaxies of different morphological type. We
compare these results with those estimated for the MFs of mock samples
drawn from N-body simulations. 

Let us summarize our results according to the three primary goals
mentioned in Introduction.

1) Primordial Gaussianity: The good match between the observed MFs and
the mock predictions based on the LCDM model with the initial
random--Gaussianity might be interpreted to imply that the primordial
Gaussianity is confirmed.  A more conservative interpretation is that,
given the size of the estimated uncertainties, these data do not
provide evidence for initial non--Gaussianity, i.e., the data are {\it
consistent} with primordial Gaussianity.  Unfortunately due to the
statistical limitation of current SDSS data, it is not easy to put
more quantitative statement concerning the initial Gaussianity, but
this is definitely what we intend to conduct with the improved
datasets available in near future.

Moreover, in order to go further and place more quantitative
constraints on primordial Gaussianity with upcoming data, one needs a
more precise and reliable theoretical model for the MFs which properly
describes the nonlinear gravitational effect possibly as well as
galaxy biasing beyond the simple mapping on the basis of the volume
fraction.  A perturbative approach by Matsubara (1994) combined with
an extensive simulation mock sample analysis may be a promising
strategy for this purpose.
 
2) Evolution of galaxy clustering: Galaxy biasing is another source of
uncertainty for relating the observed MFs to those obtained from the
mock samples for dark matter distributions (e.g. Hikage et
al. 2001). If LCDM is the correct cosmological model, the good match
of the MFs for mock samples from the LCDM simulations to the observed
SDSS MFs may indicate that nonlinearity in the galaxy biasing is
relatively small, at least small enough that it does not significantly
affect the MFs (The MFs as a function of $\nu_{\sigma}$ remain
unchanged for the linear biasing). Furthermore, we show that the
dependence of the nonlinearity of the biasing on the luminosity and
the morphological type of galaxies is also very small. Our
observational results that the effect of the galaxy biasing on the MFs
is small is consistent with the predictions of theoretical studies
(e.g. Benson et al. 2001).

3) Fair sample hypothesis: The observed morphological fluctuations
inferred from the comparison of Region 1 and Region 2 of `Sample 12'
are indeed consistent with those exhibited in our mock catalogues.

Thus our overall conclusion is that the simulations of the LCDM model
with primordial random--Gaussianity reproduce the observed MFs within
the statistical errors.  Differences between the MFs for the mock
samples drawn from simulations of the LCDM and SCDM models arise
mainly from their power spectra. The present MFs' analysis places
constraints on the cosmological models that are consistent with
previous results using the two--point correlation function, but does
not yet exploit the full advantage of the MFs as a complementary
statistic to the conventional two--point correlation function, at
least on the level of `Sample 12'.

While we found that the dependence of the MFs on the luminosity and
the morphological type of galaxies is weak and both distributions are
comparable within the statistical error, we believe that the
systematic differences due to these galaxy properties (as suggested
for the genus statistic by Hoyle et al. 2002) will certainly be
detectable in the near future.

\bigskip

We thank Y. P. Jing for kindly providing us his N--body simulation data
which were used in generating mock samples.  T. B. acknowledges
hospitality at the University of Tokyo and financial support by the
Research Center for the Early Universe (RESCEU).  I. K. gratefully
acknowledges support from the Takenaka-Ikueikai fellowship.  Numerical
computations were carried out at ADAC (the Astronomical Data Analysis
Center) of the National Astronomical Observatory, Japan (project ID:
mys02).  This research was also supported in part by the Grants--in--Aid
from Monbu--Kagakusho and Japan Society of Promotion of Science
(12640231, 13740150, 14102004, and 1470157), and by the
Sonderforschungsbereich (SFB) 375 `Astroparticle physics' of 
Deutsche Forschungsgemeinschaft (DFG). M.S.V. acknowledges the support
of NSF grant AST 00-71201 and the John Templeton Foundation.
J.R.G. was supported by NSF grant AST 99-00772.

Funding for the creation and distribution of the SDSS Archive has been
provided by the Alfred P. Sloan Foundation, the Participating
Institutions, the National Aeronautics and Space Administration, the
National Science Foundation, the U.S. Department of Energy, the
Japanese Monbukagakusho, and the Max Planck Society. The SDSS Web site
is http://www.sdss.org/. 

The SDSS is managed by the Astrophysical Research Consortium (ARC) for
the Participating Institutions. The Participating Institutions are The
University of Chicago, Fermilab, the Institute for Advanced Study, the
Japan Participation Group, The Johns Hopkins University, Los Alamos
National Laboratory, the Max-Planck-Institute for Astronomy (MPIA),
the Max-Planck-Institute for Astrophysics (MPA), New Mexico State
University, University of Pittsburgh, Princeton University, the United
States Naval Observatory, and the University of Washington.


\clearpage

\section*{Appendix A: Calculating higher-order Minkowski Functionals 
from the Koenderink invariants of a scalar field}

As stated in the main text, the higher-order Minkowski Functionals
$V_\mu$, $\mu=1,\ldots,d$ of any pattern $P\subset\Omega$ in $d$
dimensions can be calculated by integrating the partial Minkowski
Functionals $v_\mu({\bf x})$ along the surface of the pattern $\partial{P}$:
\begin{equation}
  V_\mu=\int_{\partial{P}}{\mathrm d}^{d-1}S\,v_\mu({\bf x}),
\end{equation}
where ${\mathrm d}^{d-1}S$ is the $d{-}1$-dimensional surface element at
${\bf x}$.  Unfortunately, this formula is of little practical use for
evaluating the Minkowski Functionals of a pattern $P_\nu$ given as the
isocontour of a scalar field $u({\bf x})$, that is
\begin{equation}
  P_\nu=\left\{ {\bf x} | u({\bf x})\ge\nu \right\}.
\end{equation}
We aim to convert the surface integrals above into volume integrals of
functions of the field and its derivatives.

It is relatively simple to convert the surface integral itself; we
have
\begin{equation}
  \int_{\partial{P_\nu}}{\mathrm d}^{d-1}S
  =
  \int_\Omega{\mathrm d}^dx\,|\nabla{u({\bf x})}|\delta(u({\bf x})-\nu),
\end{equation}
where $\nabla{u({\bf x})}$ is the gradient of the field and
$\delta(u({\bf x})-\nu)$ is a Dirac $\delta$-function selecting only the
surface of the pattern from the whole support $\Omega$.

As far as the local Minkowski Functionals $v_\mu({\bf x})$ are concerned,
we will restrict ourselves to the relevant case of three dimensions.
Then, the local Minkowski Functionals are proportional to the mean and
Gaussian curvatures, $H$ and $G$, of the surface:
\begin{eqnarray}
  v_1({\bf x}) &=& \frac{1}{6} \\
  v_2({\bf x}) &=& \frac{1}{3\pi}H \\
  v_1({\bf x}) &=& \frac{1}{4\pi}G.
\end{eqnarray}
In order to express the curvatures in terms of derivatives, we perform
some textbook differential geometry.  The surface around a point ${\bf x}$
can be parameterized locally with two parameters $t_a$, $a=1,2$.  The
isocontour of the field $u({\bf x})$ at threshold $\nu$ is characterized
by a simple implicit equation:
\begin{equation}
  \nu = u({\bf x}).
\end{equation}
Taking the derivative of this equation with respect to the surface
parameters $t_a$ yields a system of equations that allows us to
determine the fundamental forms of the iso--surface.  We
have\footnote{In the following, ordinary partial derivatives with
respect to a coordinate in three dimensions are denoted by the index
of the coordinate following a comma, while derivatives along the
surface are denoted with the index of the surface parameter following
a semicolon.  Indices named $a$ and $b$ are used for the surface
parameters and therefore run over $1,2$, while indices named
$i,j,\ldots$ run from 1 to 3.  Summation over indices occurring in
pairs is understood.}:
\begin{equation}
  0=u_{;a}=u_{,i}x_{i;a},
  \qquad{\rm and}\qquad
  0=u_{;ab}=u_{,ij}x_{i;a}x_{j;b}+u_{,i}x_{i;ab}.
  \label{iso}
\end{equation}

The first part of the system,
\begin{equation}
  0=u_{,i}x_{i;a},
\end{equation}
can be solved to yield the tangent vectors ${\bf x}_{;a}$, $a=1,2$ of the
surface.  We will use\footnote{In fact, this solution is not unique,
reflecting the freedom of parameterization.  However, as long as the
tangent vectors are not colinear, the final results, namely the local
Minkowski Functionals, are independent of the parameterization.}
\begin{equation}
  x_{i;a}=\varepsilon_{aij}u_{,j}.
\end{equation}
The scalar products of the tangent vectors form the entries of the
first fundamental form of the surface, the metric tensor $g_{ab}$.  We
have
\begin{equation}
  g_{ab}
  =
  x_{i;a}x_{i;b}
  =
  \delta_{ab}u_{,i}u_{,i}-u_{,a}u_{,b}.
\end{equation}
In the following, we need the determinant $g$ of the
metric tensor, given as
\begin{equation}
  g = \det g_{ab} = u_{,3}^2u_{,i}u_{,i}
\end{equation}
and its inverse
\begin{equation}
  g^{ab} = \frac{\delta_{ab}u_{,3}^2+u_{,a}u_{,b}}{g}.
\end{equation}

In order to obtain the second fundamental form of the surface, we need
the second derivatives $x_{i;ab}$ of the surface and the components
$n_i$ of the normal vector.  Equation~(\ref{iso}) yields the condition
\begin{equation}
  u_{,i} x_{i;ab} = - u_{,ij} x_{i;a} x_{j;b}
\end{equation}
for the second derivatives, which will turn out to be enough for our
purposes.  The normal vector is proportional to the gradient, and has
to be normalized of course, so we choose
\begin{equation}
  n_i = -\frac{u_{,i}}{(u_{,j}u_{,j})^{1/2}}.
\end{equation}
The minus sign directs the normal vector towards regions of lower
density; this corresponds to the intuitive expectation that the maxima
of the field should be surrounded by islands of isocontour.  Finally,
we obtain the pseudo--tensor $b_{ab}$, containing the components of the
second fundamental form of the surface
\begin{equation}
  b_{ab}
  =
  - n_i x_{i;ab}
  =
  \frac{u_{,i}\varepsilon_{aij}u_{,jk}\varepsilon_{bkl}u_{,l}}
       {(u_{,m}u_{,m})^{1/2}}.
\end{equation}
Again, we need the determinant $b$ of this object.  Some
algebra leads to the result
\begin{equation}
  b
  =
  \det b_{\mu\nu}
  =
  u_{,3}^2
  \frac{\varepsilon_{ijk}\varepsilon_{lmn}u_{,i}u_{,l}u_{,jm}u_{,kn}}
       {u_{,o}u_{,o}^{1/2}}.
\end{equation}

With these preparations, the mean curvature $H$ and the Gaussian
curvature $G$ of the isocontour of a scalar field can be expressed in
terms of the field's derivatives.  In terms of the components of the
first and second fundamental form, they are given as
\begin{equation}
  H=\frac{1}{2}g^{\mu\nu}b_{\mu\nu} ,\quad
  G=\frac{b}{g}.
\end{equation}
Putting the results from above into these formulae yields
\begin{eqnarray}
  H 
  &=& 
  \frac{\varepsilon_{ijk}\varepsilon_{ilm}u_{,j}u_{,l}u_{,km}}
       {2(u_{,n}u_{,n})^{3/2}},
  \\ 
  G 
  &=&
  \frac{\varepsilon_{ijk}\varepsilon_{lmn}u_{,i}u_{,l}u_{,jm}u_{,kn}}
       {(u_{,o}u_{,o})^2}.
\end{eqnarray}

\section*{Appendix B: Analysis of Magnitude--limited Samples}

As discussed above, in this paper we focus on the analysis of
volume--limited samples to avoid the systematic effects of variation
with redshift of the mean galaxy density and range of galaxy luminosity.
However, the statistical error of the MFs becomes large in
volume--limited samples due to
the small number density or the small volumes of such samples.
In apparent--magnitude limited samples we can make use of nearly all
of the data. 
Our results for volume--limited samples
suggest that the luminosity bias is small and, therefore, that it
might be reasonable to analyze
apparent--magnitude limited samples. To do so requires correction for
the variation with redshift in the expected galaxy density.

\begin{figure*}[tph]
\begin{center}
\FigureFile(80mm,80mm){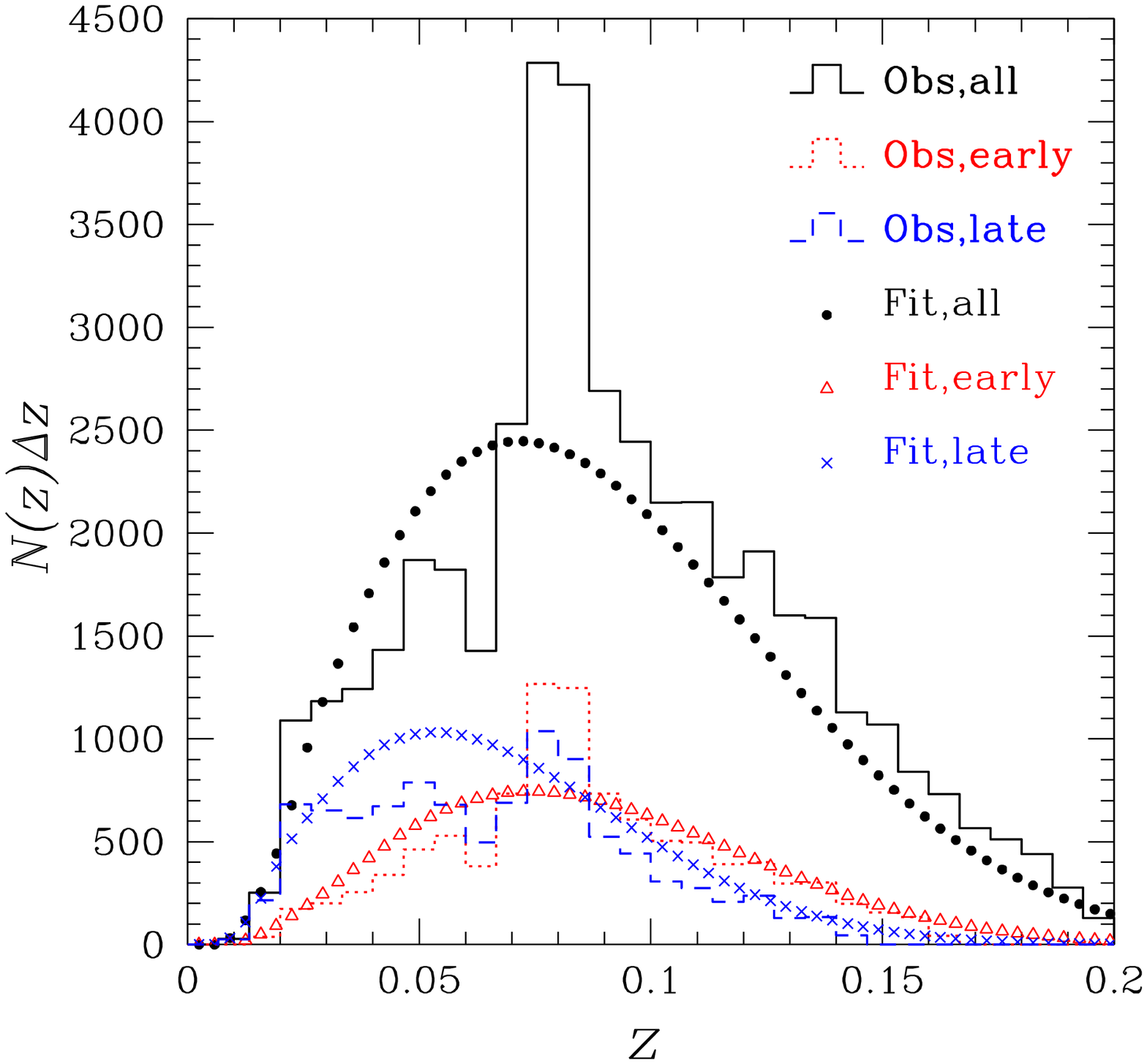}
\FigureFile(80mm,80mm){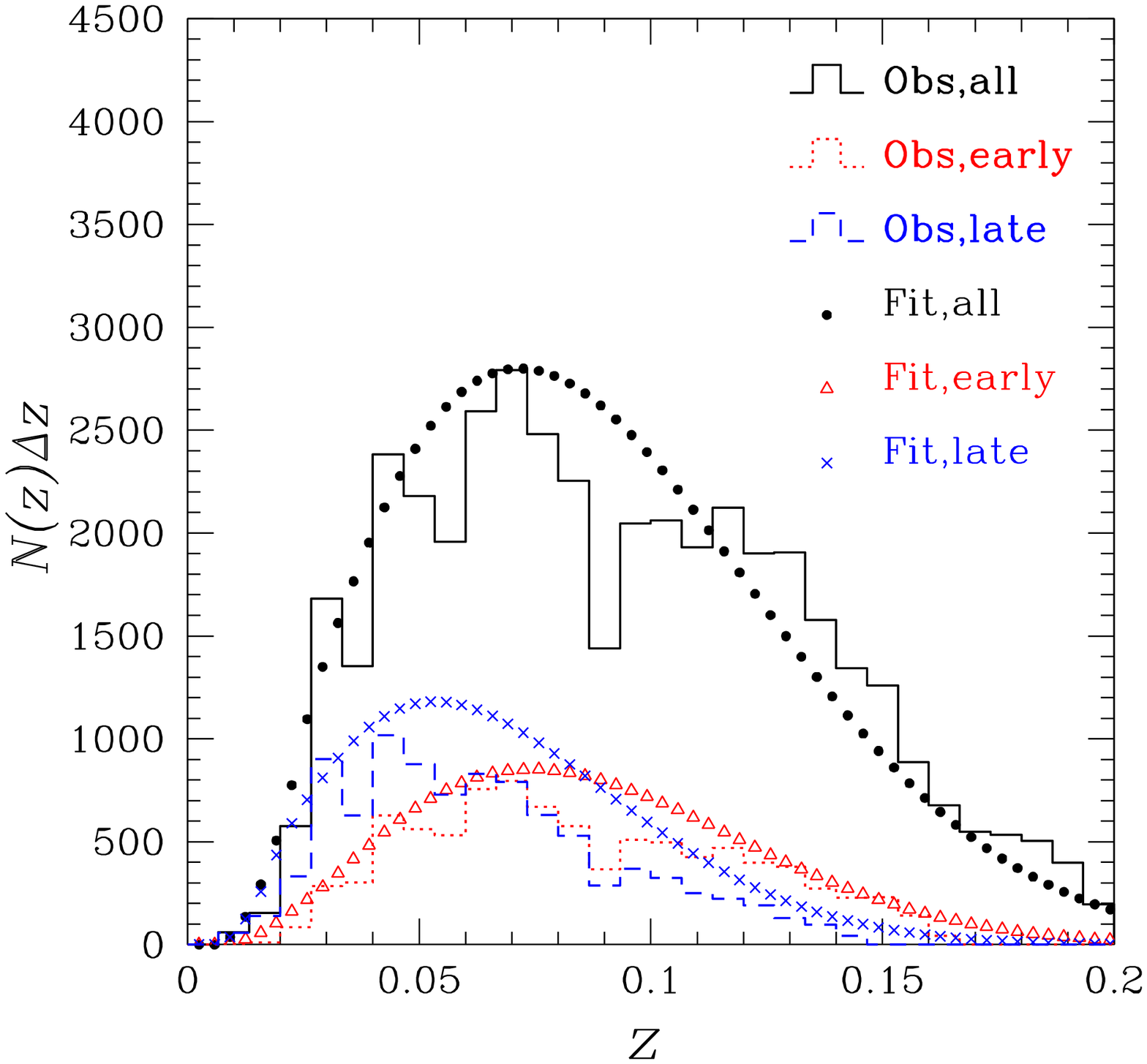} 
\caption{Number density distribution of our SDSS galaxy sample for All
(solid histogram), Early---type (dotted histogram), and Late--type
galaxies (dashed histogram) in Region 1 and Region 2, respectively.  The
estimations with the Schechter form of the luminosity function
(Eq. (\ref{eq:phiz})) fitted to the observation are also plotted for All
(filled circles), Early--type (open triangles) and Late--type (crosses)
galaxies.  The redshift width ($\Delta z$) of the histogram is $1/150$.}
\end{center}
\label{fig:Nz}
\end{figure*}

We compute the selection function $\phi(z)$, by integrating over the
Schechter form of the luminosity function,
\begin{eqnarray}
\label{eq:phiz}
\phi(z)
\propto \int^{M_{\rm max}(z)}_{M_{\rm min}(z)}
&&\phi_\ast[10^{0.4(M_\ast-M)}]^{\alpha+1} \cr
&\times& \exp[-10^{0.4(M_\ast-M)}]dM \;,
\end{eqnarray}
with parameters for all galaxies measured in the ${}^{0.1}r$ band by
Blanton et al. (2002) and those for Early-- and Late--type galaxies
measured in the $r$--band by Nakamura et al (listed in Table
\ref{tab:LF}). Note that the ${}^{0.1}r$ band corresponds to the
$r$--band shifted to match their rest--frame shape at $z=0.1$.
The limits of integration are the absolute--magnitude limits.
\begin{eqnarray}
M_{\rm max/min}(z) &=& m_{\rm max/min} \cr
&-&5\log\left[\frac{(1+z)d_c(z)}{10{\rm pc}}\right]
-K(z) .
\end{eqnarray}
We use apparent--magnitude limits $m_{\rm max}=17.5$, $m_{\rm
min}=14.5$, and apply an approximate K--correction factor $K(z) = 0.9z$
($K(z) = 0.9(z-0.1)$ for ${}^{0.1}r$ band).  Figure \ref{fig:Nz} shows
the number density distributions of our samples of All, Early--type and
Late--type galaxies (histograms) compared with fits from Equation
(\ref{eq:phiz}) for both Region 1 and Region 2.
\begin{table*}[h]
\caption{Parameter values of the Schechter form of the luminosity
function (Eq.(\ref{eq:phiz})) for All, Early--type, and Late--type
galaxies, where the ${}^{0.1}r$--band corresponds to the $r$--band
shifted to match their rest--frame shape at $z=0.1$.}
\begin{center}
\begin{tabular}{ccccc}
\hline\hline
galaxy type & band & $M_\ast - 5{\rm log}_{10}h$ & $\alpha$ 
& $\phi_\ast (10^{-2}(h^{-1}{\rm Mpc})^{-3})$ \\ \hline
All  & ${}^{0.1}r$ & $-20.44$ & $-1.05$ & $1.49$ \\
Early--type & $r$ & $-20.62$ & $-0.68$ & $0.67$ \\
Late--type & $r$ & $-20.35$ & $-1.12$ & $1.09$ \\ \hline
\end{tabular}
\end{center}
\label{tab:LF}
\end{table*}
We list the properties of the magnitude--limited samples with
absolute--magnitude range from $-23$ to $-17$ in Table \ref{tab:maglim}.
The redshift range for each sample is determined to be the region where
the mean separation of galaxies, $\phi(z)^{-1/3}$, is smaller than the
smoothing scale.  To construct mock samples with the same redshift distribution, we randomly select dark matter particles in the wedge
samples in real space according to the selection function to reproduce
the number of observed SDSS galaxies in the current samples.
To correct the galaxy density field for the variation with distance of
the selection function, we weight each galaxy by the
inverse of the selection function $\phi(z)^{-1}$ in redshift space
(e.g., Rhoads et al. 1994; Vogeley et al. 1994).
\begin{table*}[h]
\caption{Properties of each magnitude--limited sample of SDSS galaxies
including the galaxy type, the smoothing scale $R_{\rm G}$, the redshift
range, the number of SDSS galaxies in each sample $N_{\rm gal}$, and the
mean number of mass particles in six mock samples $N_{\rm particle}$.}
\begin{center}
\begin{tabular}{ccccccc}
\hline\hline
& & & \multicolumn{2}{c}{$N_{\rm gal}$} 
& \multicolumn{2}{c}{$N_{\rm particle}$} \\ 
\raisebox{1.5ex}[0pt]{type} & \raisebox{1.5ex}[0pt]{$R_{\rm G}$} & 
\raisebox{1.5ex}[0pt]{redshift range} & Region1 & Region2 & Region1 & Region2  
\\ \hline
All   & $ 5h^{-1}$Mpc & $0.009<z<0.088$ &  $21898$ & $20768$ 
&  $18325$ & $21454$   \\
All   & $ 7h^{-1}$Mpc & $0.008<z<0.121$ &  $32714$ & $30231$ 
&  $29243$ & $32950$   \\
All   & $10h^{-1}$Mpc & $0.007<z<0.150$ &  $39416$ & $37512$ 
&  $34803$ & $39689$   \\
All   & $20h^{-1}$Mpc & $0.006<z<0.197$ &  $43355$ & $41805$ 
&  $38171$ & $42050$   \\
Early & $ 7h^{-1}$Mpc & $0.013<z<0.082$ &  $ 4657$ & $ 4765$ 
&  $ 4672$ & $ 5309$   \\
Early & $10h^{-1}$Mpc & $0.009<z<0.117$ &  $ 8069$ & $ 7146$ 
&  $ 8059$ & $ 9232$   \\
Early & $20h^{-1}$Mpc & $0.007<z<0.167$ &  $ 9889$ & $ 9168$ 
&  $10153$ & $11157$   \\
Late  & $ 5h^{-1}$Mpc & $0.011<z<0.049$ &  $ 4158$ & $ 4390$ 
&  $ 4555$ & $ 5532$   \\
Late  & $ 7h^{-1}$Mpc & $0.008<z<0.078$ &  $ 7003$ & $ 7199$ 
&  $ 8262$ & $ 9038$   \\
Late  & $10h^{-1}$Mpc & $0.007<z<0.104$ &  $ 8734$ & $ 8444$ 
&  $10415$ & $11852$   \\
Late  & $20h^{-1}$Mpc & $0.006<z<0.145$ &  $ 9761$ & $ 9372$ 
&  $11891$ & $13077$  
\\ \hline
\end{tabular}
\end{center}
\label{tab:maglim}
\end{table*}
\begin{table*}[h]
\caption{Properties of the smoothed density field of each 
magnitude--limited sample including $R_{\rm G}$, 
the mesh size of the simulation box, the galaxy type, $N_{\rm res}$
(Eq. (\ref{eq:nres})), the r.m.s. fluctuation $\sigma$ of the SDSS 
galaxy number density field,
and the averaged $\sigma$ with one--sigma error 
of the particle number density field in the mock samples.}
\begin{center}
\begin{tabular}{cccccccc}
\hline\hline
$R_{\rm G}$ & & \multicolumn{2}{c}{$N_{\rm res}$(volume fraction)} 
&  \multicolumn{2}{c}{$\sigma$ of SDSS galaxies} 
&  \multicolumn{2}{c}{$\sigma$ of mock samples} \\ 
(mesh size) & \raisebox{1.5ex}[0pt]{type} & Region1 & Region2 
& Region1 & Region2 & Region1 & Region2 
\\ \hline
$ 5h^{-1}$Mpc     & All   & $464(0.71)$ & $486(0.65)$ & $1.17$ 
& $0.97$ & $1.07\pm0.13$ & $0.97\pm0.10$\\
($2.1h^{-1}$Mpc)  & Late  & $112(0.59)$ & $112(0.52)$ & $0.86$ 
& $0.74$ & $0.99\pm0.12$ & $0.94\pm0.19$\\
$ 7h^{-1}$Mpc     & All   & $424(0.70)$ & $441(0.64)$ & $0.86$ 
& $0.88$ & $0.83\pm0.09$ & $0.80\pm0.07$\\
($2.9h^{-1}$Mpc)  & Early & $113(0.59)$ & $114(0.52)$ & $0.98$ 
& $0.74$ & $0.84\pm0.12$ & $0.80\pm0.10$\\
                  & Late  & $121(0.59)$ & $122(0.52)$ & $0.66$ 
& $0.64$ & $0.86\pm0.14$ & $0.81\pm0.11$\\
$10h^{-1}$Mpc     & All   & $182(0.46)$ & $177(0.39)$ & $0.63$ 
& $0.64$ & $0.66\pm0.05$ & $0.66\pm0.04$\\
($3.8h^{-1}$Mpc)  & Early & $ 65(0.36)$ & $ 62(0.30)$ & $0.73$ 
& $0.72$ & $0.63\pm0.07$ & $0.59\pm0.05$\\
                  & Late  & $ 48(0.33)$ & $ 45(0.27)$ & $0.66$ 
& $0.55$ & $0.59\pm0.07$ & $0.59\pm0.06$\\
$20h^{-1}$Mpc     & All   & $ 29(0.27)$ & $ 26(0.21)$ & $0.34$ 
& $0.45$ & $0.33\pm0.05$ & $0.35\pm0.04$\\
($5.3h^{-1}$Mpc)  & Early & $ 12(0.20)$ & $ 10(0.14)$ & $0.34$ 
& $0.43$ & $0.32\pm0.07$ & $0.31\pm0.06$\\
                  & Late  & $  7(0.15)$ & $  5(0.11)$ & $0.28$ 
& $0.30$ & $0.31\pm0.07$ & $0.32\pm0.07$
\\ \hline
\end{tabular}
\end{center}
\label{tab:maglim2}
\end{table*}
Table \ref{tab:maglim2} lists $N_{\rm res}$ (Eq. (\ref{eq:nres})) and
$\sigma$ of galaxies in each sample and $\sigma$ with one--sigma error
estimated from the mock samples.  For all of the smoothing lengths the
table shows the general feature that clustering is strong for
Early--type galaxies, and weak for Late--type galaxies.

\begin{figure*}[tph]
\begin{center}
\FigureFile(95mm,95mm){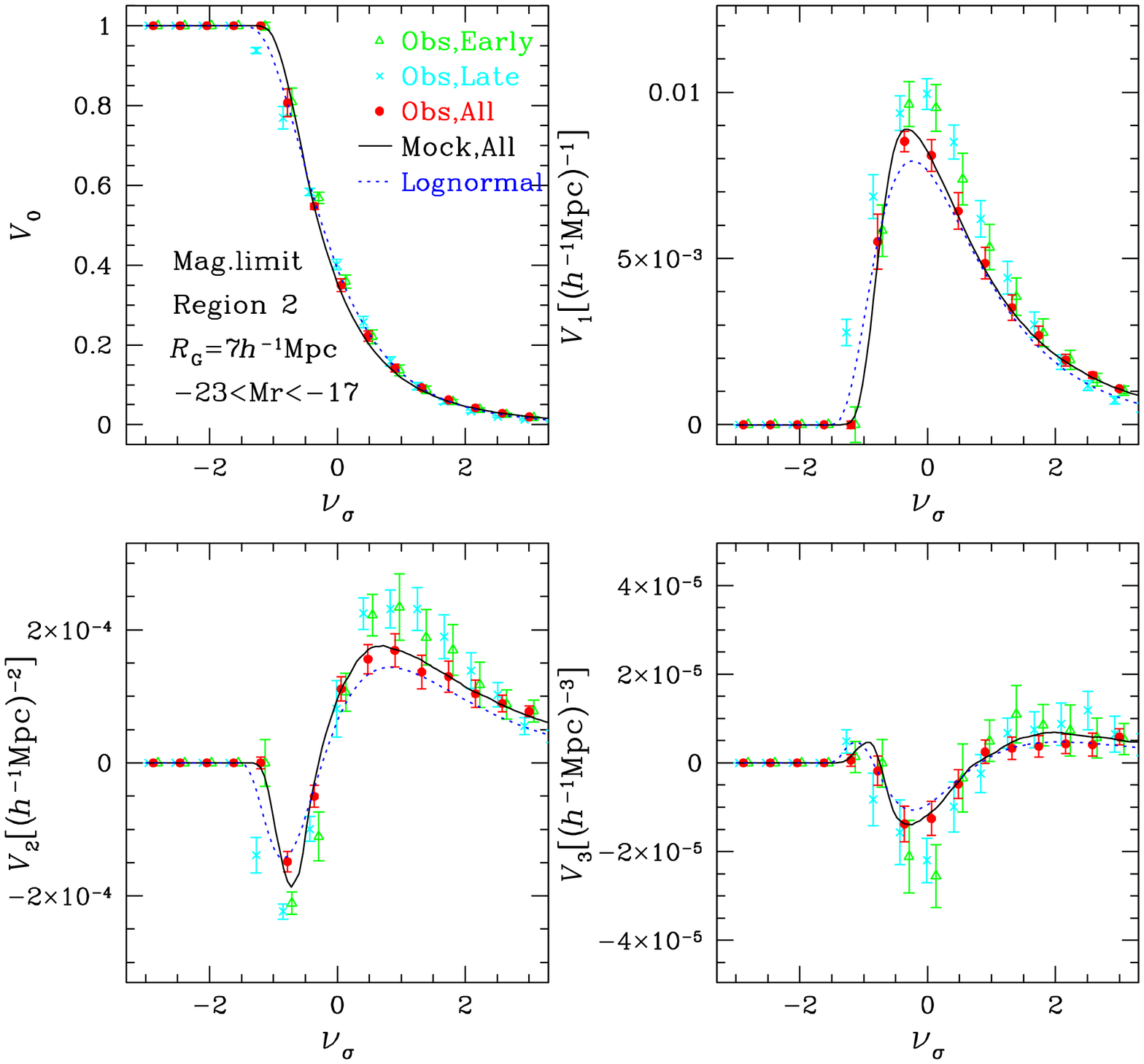}
\FigureFile(95mm,95mm){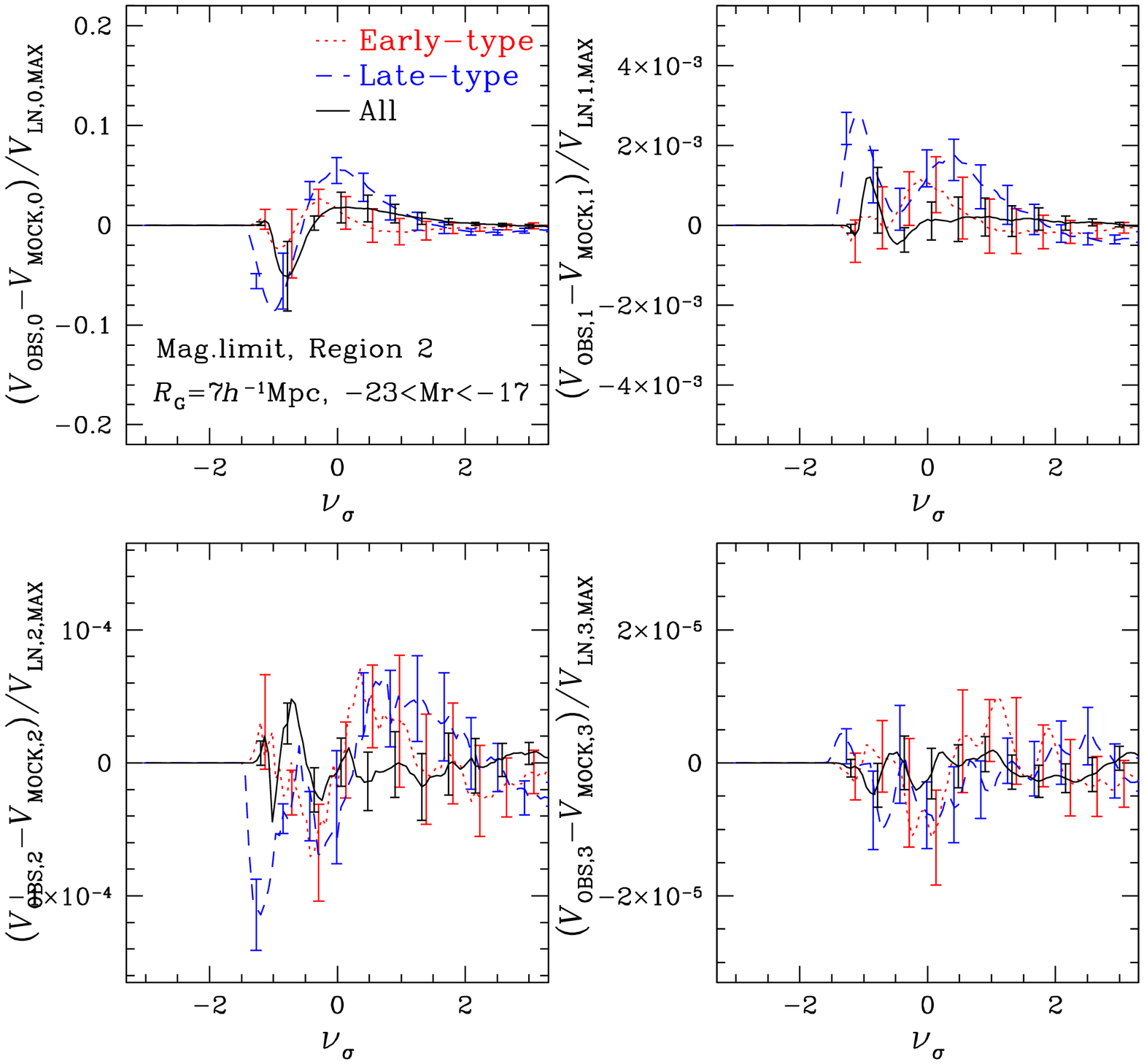}
\caption{ {\it Upper:} Comparison of the MFs between the
magnitude--limited samples for All (filled circles), Early--type (open
triangles) and Late--type galaxies (crosses).  Error bars, estimated
from the mock results, are added to the observational results. The
averaged MFs from mock samples for all galaxies are also plotted with
solid lines.  The log--normal model predictions (Eq. (\ref{eq:log-normal}))
are plotted with dotted lines. {\it Lower:} Dependency of the galaxy
type on the MFs for magnitude--limited samples of SDSS
galaxies. Subtractions of the MFs at each $\nu_{\sigma}$ of the mock results
from the observed MFs are plotted for Early--type (dotted lines),
Late--type (dashed lines) and All (solid lines) galaxies with error bars
estimated from the mock results. We adopt the smoothing scale of $R_{\rm
G}=7h^{-1}$Mpc and the data of Region 2.}
\end{center}
\label{fig:mag_type}
\end{figure*}
Figure \ref{fig:mag_type} shows the comparison of MFs for All,
Early---type and Late--type galaxies with error bars estimated from mock
samples for each type of galaxies.  The averaged MFs of mock samples for
all galaxies in the LCDM model and the log--normal predictions are also
plotted.  Focusing on the difference in morphological properties of the
distribution due to the morphological type of galaxies, we plot the
subtraction of the simulated results from the observed MFs for each type
of galaxies, normalized by the MFs predicted from the log--normal model,
in the lower panel of Figure \ref{fig:mag_type}.  We conclude that also
in the magnitude--limited samples the difference due to morphological
type of galaxies cannot be appreciated, as was found for the
volume--limited samples.

\end{document}